\documentclass[aps,prd,twocolumn,superscriptaddress,nofootinbib,
               floatfix,longbibliography]{revtex4-2}

\usepackage{amsmath,amssymb,amsfonts}
\usepackage{graphicx}
\usepackage{bm}
\usepackage{xcolor}
\usepackage[colorlinks=true,linkcolor=blue,citecolor=blue,urlcolor=blue]{hyperref}

\newcommand{\ed}[1]{#1}

\newcommand{\at}{\tilde{\alpha}}
\newcommand{\rs}{R_{\mathrm{stop}}}
\newcommand{\ratf}{r_{0}^{\mathrm{atf}}}
\newcommand{\rrtf}{r_{0}^{\mathrm{rtf}}}
\newcommand{\Krr}{K^{\hat r}{}_{\hat r}}
\newcommand{\Kii}{K^{\hat\imath}{}_{\hat\imath}}

\begin{document}

\title{Tidal forces in the quantum Oppenheimer--Snyder black hole}

\author{Anuar Idrissov}
\email{anuar.idrissov@gmail.com}
\affiliation{Instituto de Ciencias Nucleares, Universidad Nacional Aut\'onoma de M\'exico, Mexico}
\affiliation{Fesenkov Astrophysical Institute, Observatory 23, 050020, Almaty, Kazakhstan}
\affiliation{Al-Farabi Kazakh National University, Al-Farabi Ave.\ 71, 050040, Almaty, Kazakhstan}

\author{Hernando~\surname{Quevedo}}
\email[]{quevedo@nucleares.unam.mx}
\affiliation{Instituto de Ciencias Nucleares, Universidad Nacional Aut\'onoma de M\'exico, Mexico}
\affiliation{Dipartimento di Fisica and ICRA, Universit\`a di Roma ``La Sapienza'', Roma, Italy}
\affiliation{Al-Farabi Kazakh National University, Al-Farabi Ave.\ 71, 050040, Almaty, Kazakhstan}

\date{\today}

\begin{abstract}
We investigate the tidal forces experienced by massive particles in radial free
fall in the quantum Oppenheimer--Snyder black hole, a loop quantum gravity
correction to the Schwarzschild geometry that describes the exterior of a
bouncing dust ball. Using an orthonormal
tetrad adapted to a freely falling observer, we show that the radial and
angular components of the tidal tensor reproduce their Schwarzschild
counterparts at large distances but reverse sign in the interior. The angular
zero remains confined between the Cauchy and event horizons and coincides with
the latter only in the extremal limit. A particle released from rest does not
reach the center, it stops at a turning point inside the Cauchy horizon, where
the ratio of the tidal components is independent of the parameters of the
solution and is fixed by the transverse equation of state of the effective
source. Solving the geodesic deviation
equations for two sets of initial conditions, we find that the radial deviation
vector attains its maximum at the minimum of the metric function, exactly for
one set of initial conditions and asymptotically for the other, and remains
finite throughout, in contrast with the Schwarzschild case, in which it diverges at
the singularity. This regularity originates in the bounce rather than in a
regular core, and therefore protects timelike radial infall while leaving
radial null geodesics unaffected. Examining the full range of the quantum
parameter, we find that its sign \ed{determines} whether the tidal sector possesses any
of this structure, while its magnitude \ed{determines} only whether that structure is
hidden: beyond the extremal value the geometry becomes horizonless, and the
entire deformation history, bounce included, is exposed to distant observers.
\end{abstract}

\maketitle

\section{Introduction}
\label{sec:intro}

Black-hole spacetimes provide a direct setting in which to test general
relativity in the strong-field regime. Horizon-scale observations of M87* and
Sagittarius A* by the Event Horizon Telescope Collaboration
\cite{EHT2019,EHT2022}, together with the detection of gravitational waves from
compact-binary mergers \cite{LIGO2016}, have made this regime accessible to
observation and strengthened the motivation for studying departures from the
classical vacuum solutions of Einstein's equations. These solutions are
geodesically incomplete: the Schwarzschild geometry \cite{Schwarzschild1916}
contains a curvature singularity at $r=0$, while the singularity theorems
\cite{Penrose1965,HawkingEllis} establish the generic nature of geodesic
incompleteness under suitable conditions.

Loop quantum gravity provides a framework in which such singular
\ed{behavior} can be modified by quantum geometry. In loop quantum cosmology,
the big-bang singularity is replaced by a quantum bounce when the energy
density reaches the Planck scale \cite{APS2006}. Applying the corresponding
effective dynamics to the Oppenheimer--Snyder collapse of a homogeneous dust
ball, Lewandowski, Ma, Yang and Zhang \cite{LMYZ2023} obtained a collapsing
solution that reaches a maximum density and then re-expands. Its exterior
geometry is described by
\begin{equation}
f(r)=1-\frac{2M}{r}+\frac{\alpha M^{2}}{r^{4}},
\label{eq:fintro}
\end{equation}
where $\alpha=16\sqrt{3}\pi\gamma^{3}\ell_{\mathrm{Pl}}^{2}$ and $\gamma$ is the
Barbero--Immirzi parameter. This geometry is referred to as the quantum
Oppenheimer--Snyder (qOS) black hole. It approaches Schwarzschild at large
radius and has two horizons for sufficiently small $\alpha/M^{2}$. At small
radius, however, the correction term dominates and $f(r)\to+\infty$ as
$r\to0$.

The qOS geometry has been studied in the context of shadows and perturbative
stability \cite{YZM2023}, \ed{shadows and photon rings} \cite{Ye2024},
\ed{circular orbits and thin accretion disks} \cite{ShuHuang2025}, quasinormal
modes \cite{Skvortsova2024,Gong2024}, higher-dimensional extensions
\cite{Shi2024,Jiang2026}, charged extensions \cite{Mazhari2025,Mazhari2026},
\ed{Lorentz-term} extensions \cite{OuZhang2025}, and extreme-mass-ratio
inspirals \cite{Fu2025,Yang2025EMRI,Zhang2026EMRI}. These analyses probe the
geometry primarily through global or asymptotic observables, such as orbital
frequencies, scattering properties, quasinormal spectra, or photon-ring
structure.

Tidal forces provide a local probe of the spacetime curvature. They are
determined by the Riemann tensor and can be obtained from the geodesic
deviation equation in a frame comoving with a freely falling observer. Tidal
effects have been studied in Reissner--Nordstr\"om
\cite{CrispinoRN}, the $q$-metric \cite{Idrissov2025}, charged Hayward
spacetimes \cite{HaywardTidal}, dirty black holes \cite{DirtyBH},
Simpson--Visser black bounces \cite{SimpsonVisser}, Kottler spacetimes
\cite{Kottler}, four-dimensional Einstein--Gauss--Bonnet gravity \cite{EGB},
generalized-uncertainty-principle corrected Schwarzschild
\cite{GUP}, black-bounce geometries \cite{BlackBounce,BlackBounceErratum},
Letelier--Alencar string clouds \cite{Letelier}, \ed{multi-horizon black holes
in nonlinear electrodynamics \cite{AfsharSadeghi2026},} and the Dymnikova
regular black hole \cite{Dymnikova2026}. In these examples, the tidal
components can change sign, leading to transitions between radial stretching
and compression. This does not occur in Schwarzschild spacetime, where the
radial and angular tidal components retain fixed signs.

To the best of our knowledge, the tidal sector of the qOS geometry has not
previously been \ed{analyzed}. We compute the tidal tensor in a freely falling
orthonormal frame, determine the radii at which its components vanish or
become extremal, identify the turning point of radial infall, and solve the
geodesic deviation equations for two sets of initial conditions. The qOS
geometry exhibits two features that are of particular interest. First, the
$r^{-4}$ correction leads to simple analytic relations among the characteristic
radii and tidal extrema. Second, although the curvature singularity at
$r=0$ is stronger than in Schwarzschild, the tidal forces experienced by an
infalling body remain finite because the trajectory turns around before
reaching the singularity. The finiteness of the measured tidal forces is
therefore a consequence of the infalling trajectory rather than of curvature
regularity.

This paper is \ed{organized} as follows. Section~\ref{sec:qos} introduces the
qOS geometry and derives the radial geodesics and turning point.
Section~\ref{sec:tidal} constructs the freely falling tetrad and tidal tensor
and analyses its radial and angular components.
Section~\ref{sec:deviation} solves the corresponding geodesic deviation
equations. Section~\ref{sec:origin} examines the origin of the resulting tidal
features by expressing the tidal tensor in terms of an effective source and by
studying the dependence on the sign and magnitude of $\alpha$.
Section~\ref{sec:conclusion} summarizes the results. We use geometric units
$G=c=1$ and metric signature $(-,+,+,+)$.

\section{Metric and radial geodesics}
\label{sec:qos}

\subsection{The qOS geometry}
\label{sec:geometry}

The qOS exterior is static and spherically symmetric,
\begin{equation}
ds^{2}=-f(r)\,dt^{2}+\frac{dr^{2}}{f(r)}+r^{2}d\Omega^{2},
\label{eq:metric}
\end{equation}
where $d\Omega^{2}=d\theta^{2}+\sin^{2}\theta\,d\varphi^{2}$ and
\begin{equation}
f(r)=1-\frac{2M}{r}+\frac{\alpha M^{2}}{r^{4}},
\label{eq:f}
\end{equation}
where $M$ is the ADM mass and $\alpha>0$ has dimensions of length squared.
It is convenient to introduce
\begin{equation}
x\equiv\frac{r}{M},
\qquad
\at\equiv\frac{\alpha}{M^{2}},
\end{equation}
in terms of which
\begin{equation}
f(x)=1-\frac{2}{x}+\frac{\at}{x^{4}}.
\label{eq:fx}
\end{equation}
The Schwarzschild solution is recovered at $\at=0$.

For every $\at>0$, the correction dominates at sufficiently small radius and
$f(r)\to+\infty$ as $r\to0$. The metric is therefore qualitatively different
from Schwarzschild in the innermost region. At the same time, the curvature is
not regular there. The Kretschmann scalar is
\begin{equation}
\mathcal{K}=\frac{48M^{2}}{r^{6}}
-\frac{240\alpha M^{3}}{r^{9}}
+\frac{468\alpha^{2}M^{4}}{r^{12}},
\label{eq:kretschmann}
\end{equation}
and diverges as $r^{-12}$. Thus the quantum correction does not
\ed{regularize} the curvature singularity.

This \ed{behavior} is consistent with the construction of the qOS solution:
Eq.~\eqref{eq:metric} describes the exterior vacuum region outside the
collapsing dust ball and is matched to the semiclassical interior across a
timelike hypersurface \cite{LMYZ2023}. The region near $r=0$ is therefore not
part of the physical exterior construction.

\subsection{Horizons and parameter range}
\label{sec:horizons}

The horizon radii satisfy
\begin{equation}
x^{4}-2x^{3}+\at=0.
\end{equation}
Rather than solving this equation for $x$, it is useful to regard $\at$ as a
function of the horizon radius,
\begin{equation}
\at=x_{h}^{3}(2-x_{h}).
\end{equation}
This function has a single maximum at $x_{h}=3/2$, giving
$\at_{\mathrm{ext}}=27/16$. Hence the geometry has two horizons
$x_{-}<3/2<x_{+}$ for $\at<27/16$, a degenerate horizon at
$\at=27/16$, and no horizons for $\at>27/16$ \cite{LMYZ2023,YZM2023}. In
dimensional form,
\begin{equation}
M\ge M_{\mathrm{crit}}=\frac{4}{3\sqrt3}\sqrt{\alpha},
\qquad
r_{\mathrm{crit}}=\frac{2}{\sqrt3}\sqrt{\alpha}.
\label{eq:Mcrit}
\end{equation}

For small $\at$, the two roots behave as
\begin{equation}
x_{+}=2-\frac{\at}{8}+\mathcal{O}(\at^{2}),
\end{equation}
and
\begin{equation}
x_{-}
=\left(\frac{\at}{2}\right)^{1/3}
\left[1+\frac{1}{6}
\left(\frac{\at}{2}\right)^{1/3}+\cdots\right].
\end{equation}
Thus the event horizon is slightly smaller than the Schwarzschild value,
while the inner horizon approaches the origin as $\at\to0$.

The parameter range used below is
\begin{equation}
0<\at\le\frac{27}{16}.
\end{equation}
The values $\at=0.3$, $0.9$, and $1.5$ span this black-hole regime. They should
not, however, be interpreted as astrophysical values: since
$\alpha\sim\ell_{\mathrm{Pl}}^{2}$, $\at\sim1$ corresponds to a Planck-scale
mass, whereas for $M\sim M_{\odot}$ one has $\at\sim10^{-76}$.

\begin{figure}[t]
  \includegraphics[width=\columnwidth]{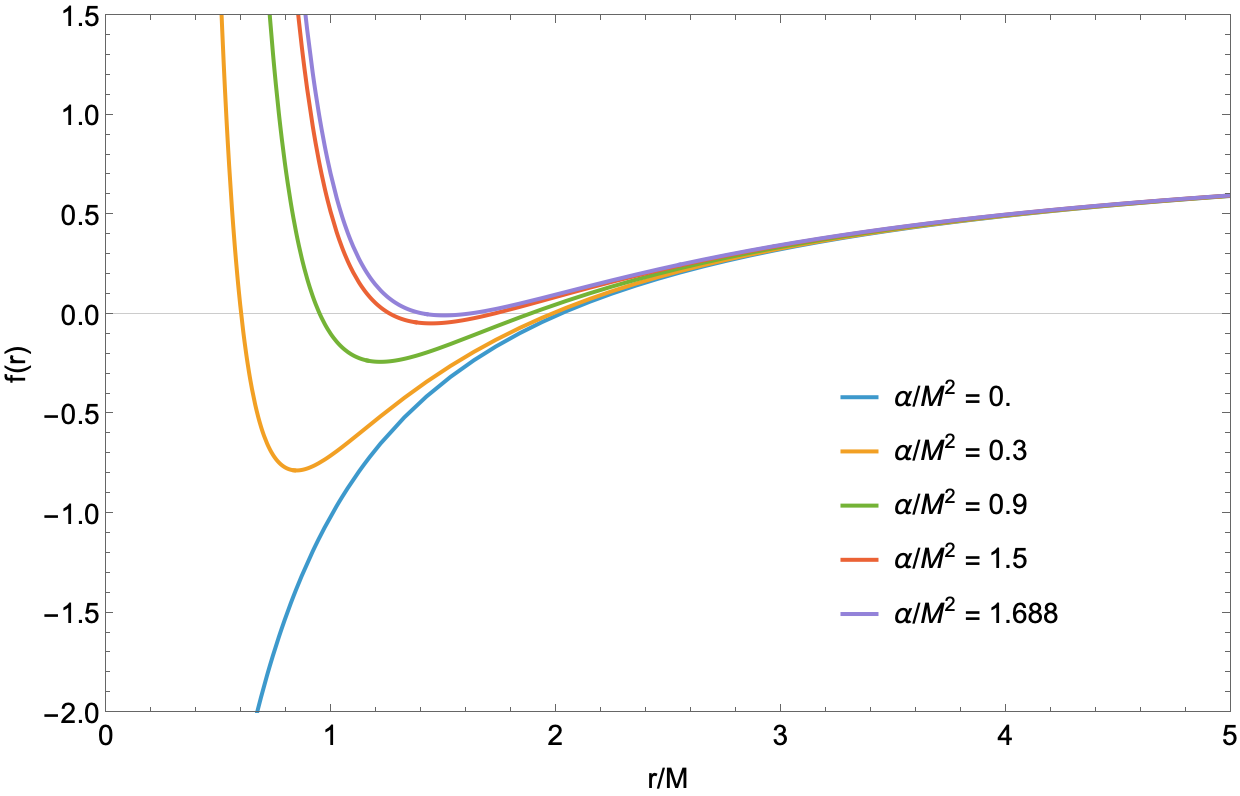}
  \caption{The metric function $f$ versus $r/M$ for several values of
    $\at=\alpha/M^{2}$; $\at=0$ corresponds to Schwarzschild. As $\at$
    increases, the two horizons approach each other and merge at
    $\at=27/16$. For every $\at>0$, $f\to+\infty$ as $r\to0$, so the
    innermost region is repulsive and a radial timelike geodesic released
    from rest outside the event horizon does not reach the \ed{center}.}
  \label{fig:f}
\end{figure}

Figure~\ref{fig:f} illustrates the corresponding \ed{behavior} of the metric
function across the parameter range considered here.

\subsection{Radial infall and the turning point}
\label{sec:geodesics}

We now consider radial timelike geodesics. For
$\dot\theta=\dot\varphi=0$, the normalization condition gives
\begin{equation}
f\dot t^{2}-f^{-1}\dot r^{2}=1,
\end{equation}
where the overdot denotes differentiation with respect to proper time $\tau$.
The conserved energy per unit mass is
\begin{equation}
E=f\dot t.
\end{equation}
For a particle released from rest at $r=b>r_{+}$, one has
$E=\sqrt{f(b)}$, and therefore
\begin{equation}
\dot r^{2}=E^{2}-f(r).
\label{eq:firstintegral}
\end{equation}
Differentiating gives
\begin{equation}
\ddot r=-\frac{1}{2}f'(r).
\label{eq:rddot}
\end{equation}
This form is regular at the horizons and is used for the numerical integration
of the geodesic deviation equations in Sec.~\ref{sec:deviation}.

Since $f(r)\to+\infty$ as $r\to0$, the right-hand side of
Eq.~\eqref{eq:firstintegral} must vanish at a finite radius. The particle
therefore reaches a turning point $r=\rs$ determined by
\begin{equation}
E^{2}=f(\rs),
\end{equation}
or, in terms of $x=r/M$,
\begin{equation}
(1-E^{2})x_{\mathrm{stop}}^{4}
-2x_{\mathrm{stop}}^{3}+\at=0.
\label{eq:turning}
\end{equation}
The radial velocity then changes sign and the particle re-expands.

For release from rest at infinity, $E=1$, and the quartic reduces to
\begin{equation}
\rs=\left(\frac{\alpha M}{2}\right)^{1/3},
\qquad
x_{\mathrm{stop}}=\left(\frac{\at}{2}\right)^{1/3}.
\label{eq:rstopinf}
\end{equation}
The result has a simple interpretation in terms of the Misner--Sharp mass.
Writing
\begin{equation}
f(r)=1-\frac{2m(r)}{r},
\qquad
m(r)=M-\frac{\alpha M^{2}}{2r^{3}},
\label{eq:misnersharp}
\end{equation}
the marginally bound equation becomes
\begin{equation}
\dot r^{2}=\frac{2m(r)}{r}.
\end{equation}
The turning point is therefore located at $m(r)=0$. For smaller radii,
$m(r)<0$, which corresponds to the repulsive inner region.

\subsection{Location of the bounce}
\label{sec:bounce}

The turning point always lies inside the Cauchy horizon,
\begin{equation}
\rs<r_{-}.
\end{equation}
This follows directly from the \ed{behavior} of $f(r)$. Since the particle is
released from rest at $b>r_{+}$, one has $f(b)>0$, and hence
$f(\rs)=f(b)>0$. The particle crosses the event horizon before reaching the
turning point, so $\rs<r_{+}$. Between the two horizons, however, $f<0$.
The only region with $r<r_{+}$ in which $f$ is positive is therefore
$r<r_{-}$.

The turning point moves inward as the release radius increases. In the two
limits,
\begin{equation}
\rs\to r_{-}\ \ (b\to r_{+}),
\qquad
\rs\to\Big(\frac{\alpha M}{2}\Big)^{1/3}\ \ (b\to\infty).
\end{equation}
Thus $\rs$ lies in the interval
\begin{equation}
\left(\frac{\alpha M}{2}\right)^{1/3}<\rs<r_{-}.
\end{equation}

The same qualitative \ed{behavior} occurs in the Dymnikova
\cite{Dymnikova2026} and charged Hayward \cite{HaywardTidal} geometries. The
location of the bounce is also relevant to the expected mass-inflation
instability of the Cauchy horizon \cite{PoissonIsrael1989,PoissonIsrael1990};
we return to this point in Sec.~\ref{sec:conclusion}.

Table~\ref{tab:radii} lists $\rs$ together with the
tidal radii derived below; already at $b=100M$ the numerically determined value
agrees with Eq.~\eqref{eq:rstopinf} to within $0.31\%$.

\begin{table}[t]
\caption{Characteristic radii, in units of $M$, for the three values of
$\at=\alpha/M^{2}$ used throughout. $x_{\mp}$ are the Cauchy and event
horizons; $\ratf=(2\at)^{1/3}$ and $\rrtf=(5\at)^{1/3}$ are the radii at which
the angular and radial tidal forces vanish; $\rs$ is the turning point for a
particle released from rest at $b=100M$, and the last column its analytic
$b\to\infty$ value $(\at/2)^{1/3}$.}
\label{tab:radii}
\begin{ruledtabular}
\begin{tabular}{lcccccc}
$\at$ & $x_{-}$ & $x_{+}$ & $\ratf$ & $\rrtf$
      & $\rs$ & $(\at/2)^{1/3}$ \\
\hline
0.3 & 0.5981 & 1.9602 & 0.8434 & 1.1447 & 0.5323 & 0.5313 \\
0.9 & 0.9499 & 1.8602 & 1.2164 & 1.6510 & 0.7683 & 0.7663 \\
1.5 & 1.2732 & 1.6883 & 1.4422 & 1.9574 & 0.9113 & 0.9086 \\
\end{tabular}
\end{ruledtabular}
\end{table}

\section{Tidal forces}
\label{sec:tidal}

\subsection{Free-fall tetrad and the tidal tensor}
\label{sec:tetrad}

The relative acceleration of two \ed{neighboring} geodesics separated by an
infinitesimal displacement $\eta^{\hat\mu}$ obeys the geodesic deviation
equation~\cite{MTW}
\begin{equation}
  \frac{D^{2}\eta^{\hat\mu}}{D\tau^{2}}
  =K^{\hat\mu}{}_{\hat\nu}\,\eta^{\hat\nu},
  \qquad
  K^{\hat\mu}{}_{\hat\nu}
  =R^{a}{}_{bcd}\,e^{\hat\mu}_{\;a}e^{b}_{\;\hat0}e^{c}_{\;\hat0}
   e^{d}_{\;\hat\nu},
  \label{eq:deviation}
\end{equation}
where hatted indices refer to the tetrad basis. We adopt the orthonormal frame
carried by an observer in radial free fall,
\begin{align}
  e^{\mu}_{\;\hat0}&=\left(\frac{E}{f},\,-\sqrt{E^{2}-f},\,0,\,0\right), \nonumber\\
  e^{\mu}_{\;\hat1}&=\left(-\frac{\sqrt{E^{2}-f}}{f},\,E,\,0,\,0\right), \nonumber\\
  e^{\mu}_{\;\hat2}&=\left(0,\,0,\,\frac{1}{r},\,0\right), \qquad
  e^{\mu}_{\;\hat3}=\left(0,\,0,\,0,\,\frac{1}{r\sin\theta}\right),
  \label{eq:tetrad}
\end{align}
which satisfies $e^{\mu}_{\;\hat\kappa}e^{\nu}_{\;\hat\sigma}g_{\mu\nu}
=\eta_{\hat\kappa\hat\sigma}$, so that measurements in this frame are those of
a local inertial observer.

Evaluating Eq.~\eqref{eq:deviation} with the Riemann tensor of
Eq.~\eqref{eq:metric} gives a diagonal tidal tensor whose components are
independent of $E$,
\begin{equation}
  \Krr=-\frac{f''}{2},
  \qquad
  \Kii=-\frac{f'}{2r},
  \qquad \hat\imath=(\hat\theta,\hat\varphi),
  \label{eq:Kgeneral}
\end{equation}
with $K^{\hat0}{}_{\hat0}$ and all off-diagonal components vanishing
identically. That the energy drops out is a general feature of radial free fall
in static spherically symmetric spacetimes and is a useful check on the
computation. Substituting the qOS metric function gives the central result of
this section,
\begin{align}
  \frac{D^{2}\eta^{\hat r}}{D\tau^{2}}
  &=\left(\frac{2M}{r^{3}}-\frac{10\,\alpha M^{2}}{r^{6}}\right)\eta^{\hat r},
  \label{eq:Krr}\\[4pt]
  \frac{D^{2}\eta^{\hat\imath}}{D\tau^{2}}
  &=\left(-\frac{M}{r^{3}}+\frac{2\,\alpha M^{2}}{r^{6}}\right)\eta^{\hat\imath},
  \label{eq:Kaa}
\end{align}
or, in dimensionless variables, $M^{2}\Krr=2x^{-3}-10\at x^{-6}$ and
$M^{2}\Kii=-x^{-3}+2\at x^{-6}$.
\ed{These components are collected in Table~\ref{tab:summary}; they are derived
in the remainder of this section, and their effect on an extended body is
analyzed in Sec.~\ref{sec:deviation}.}

For $\alpha\to0$, or equivalently $r\gg(\alpha M)^{1/3}$,
Eqs.~\eqref{eq:Krr} and~\eqref{eq:Kaa} reduce to
\begin{equation}
  \frac{D^{2}\eta^{\hat r}}{D\tau^{2}}\to\frac{2M}{r^{3}}\eta^{\hat r},
  \qquad
  \frac{D^{2}\eta^{\hat\imath}}{D\tau^{2}}\to-\frac{M}{r^{3}}\eta^{\hat\imath},
  \label{eq:schwlimit}
\end{equation}
the familiar Schwarzschild tidal forces: radial stretching and transverse
compression, the mechanism of spaghettification. In the opposite regime
$r\to0$ the quantum terms dominate,
\begin{equation}
  \Krr\to-\frac{10\alpha M^{2}}{r^{6}},
  \qquad
  \Kii\to+\frac{2\alpha M^{2}}{r^{6}}.
  \label{eq:innerlimit}
\end{equation}
Both diverge, but with the Schwarzschild signs \emph{exchanged}: near the
\ed{center} the qOS geometry compresses radially and stretches transversally.
Each component must therefore change sign somewhere, so vanishing-tidal-force
radii are guaranteed to exist.

\subsection{Radial tidal force}
\label{sec:radialtidal}

\begin{figure}[t]
  \includegraphics[width=\columnwidth]{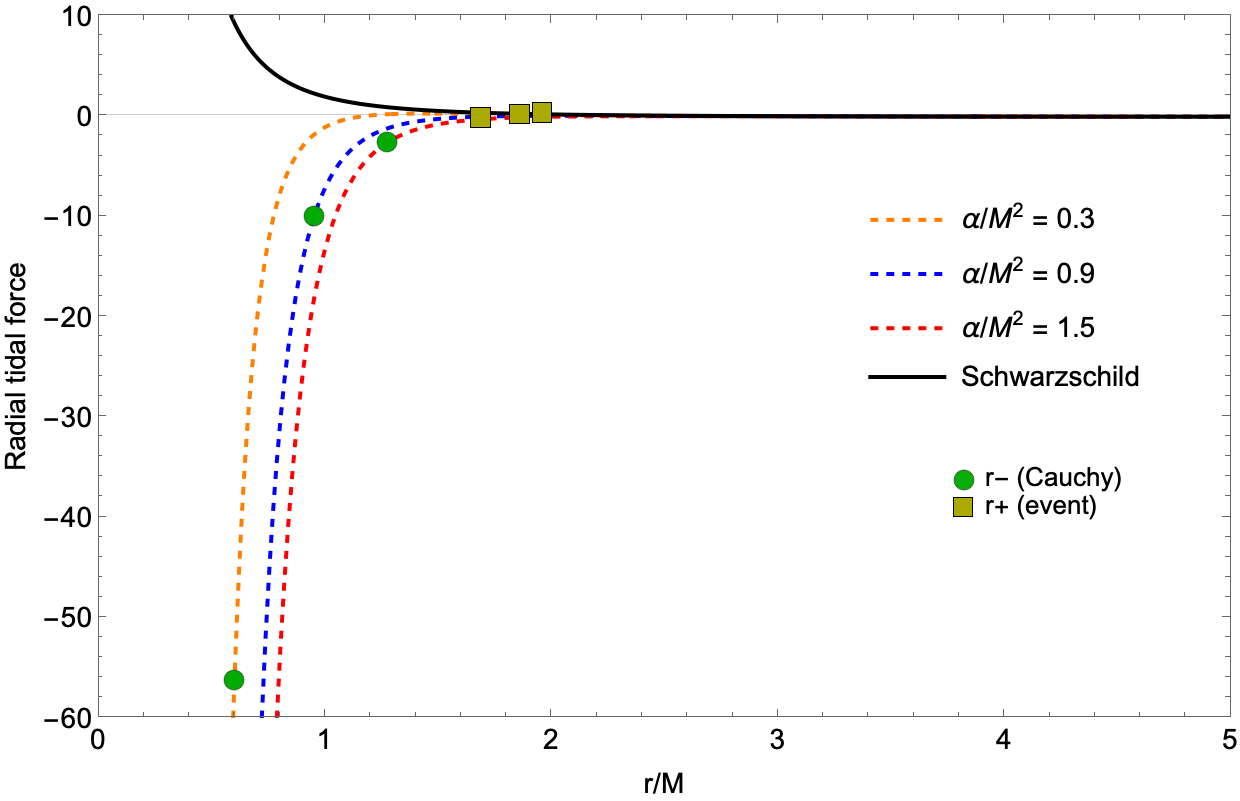}
  \caption{Radial tidal force $M^{2}\Krr$ versus $r/M$ for three values of
    $\at$, together with the Schwarzschild result (solid black), which diverges
    as $r\to0$. Green and yellow markers indicate the Cauchy horizon $r_{-}$
    and the event horizon $r_{+}$ of each curve. Each qOS curve crosses zero at
    $\rrtf=(5\alpha M)^{1/3}$, is bounded above by $1/10\at$, and \ed{decreases
    without bound} as $r\to0$: the radial stretching of the Schwarzschild
    geometry is replaced by compression in the inner region.}
  \label{fig:rtf}
\end{figure}

In Schwarzschild the radial stretching of an infalling body grows without
bound. In the qOS geometry it does not, because $\Krr$ has a single zero and a
single maximum,
\begin{equation}
  \rrtf=(5\alpha M)^{1/3},
  \quad
  \Krr\Big|_{\max}=\frac{1}{10\,\alpha}
  \ \text{at}\ r=(10\alpha M)^{1/3},
  \label{eq:rrtf}
\end{equation}
\ed{so that the radial stretching experienced by a freely falling body never
exceeds $1/10\alpha$, at any radius.} Outside $\rrtf$ the force is positive
(stretching); inside it, negative (compression). \ed{The compression is not
bounded by the geometry: by Eq.~\eqref{eq:innerlimit} it grows as $r^{-6}$.
Along a trajectory it is nevertheless bounded by the turning point, and the
range actually experienced by a body released from rest at infinity is}
\begin{equation}
  -\frac{36}{\alpha}\;\le\;\Krr\;\le\;\frac{1}{10\,\alpha},
  \label{eq:range}
\end{equation}
the lower end attained at the bounce and the upper end at
$r=(10\alpha M)^{1/3}$, one factor of $20^{1/3}$ further out.
Figure~\ref{fig:rtf} displays $M^{2}\Krr$ for several $\at$ together with the
Schwarzschild curve.

The position of this sign change relative to the horizons follows from setting
$\rrtf=r_{+}$. Combining $x^{3}=5\at$ with $x^{4}-2x^{3}+\at=0$ eliminates
$\at$ and gives $x^{4}=\tfrac{9}{5}x^{3}$, hence
\begin{equation}
  x=\frac{9}{5},
  \qquad
  \at_{\star}=\frac{x^{3}}{5}=\frac{729}{625}=1.1664.
  \label{eq:crossing}
\end{equation}
For $\at>\at_{\star}$, the radial tidal force changes sign outside the event
horizon. \ed{In this parameter range an observer falling from rest at infinity
enters a region of radial compression before crossing the horizon. In terms of
the mass, the band is $M_{\mathrm{crit}}\le M<\tfrac{25}{27}\sqrt{\alpha}
\simeq1.20\,M_{\mathrm{crit}}$, that is, masses within $20\%$ of the critical
value. The exposure increases with $\at$ and reaches
$\rrtf/r_{+}=(5/2)^{1/3}$ at extremality, where $r_{+}=\ratf$ by
Eq.~\eqref{eq:confinement} and Eq.~\eqref{eq:ratio} applies.}

\begin{figure}[t]
  \includegraphics[width=\columnwidth]{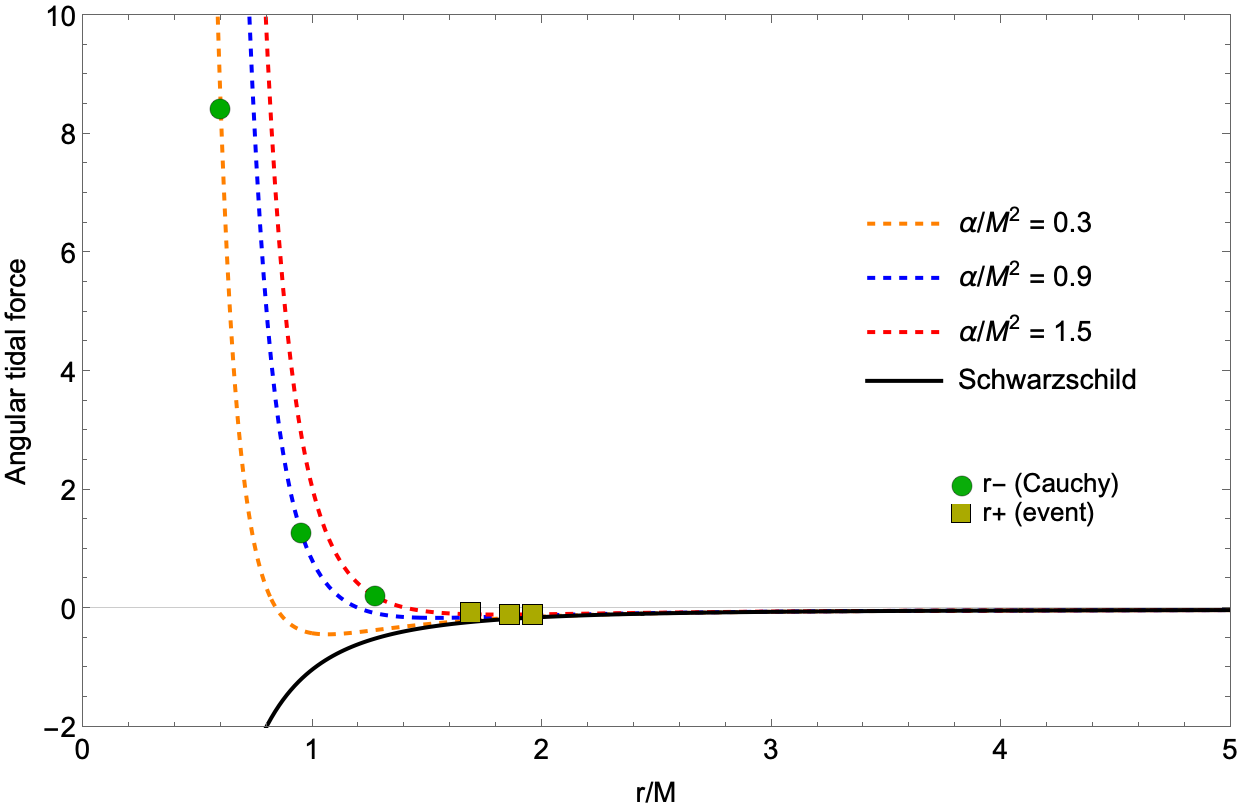}
  \caption{Angular tidal force $M^{2}\Kii$ versus $r/M$ for three values of
    $\at$, together with the Schwarzschild result (solid black). Green and
    yellow markers indicate the Cauchy and event horizons. The zero at
    $\ratf=(2\alpha M)^{1/3}$ always lies between the two horizons, touching
    $r_{+}$ only in the extremal limit $\at=27/16$.}
  \label{fig:atf}
\end{figure}

\subsection{Angular tidal force}
\label{sec:angulartidal}

The transverse sector mirrors the radial one, with its own zero and its own
bound,
\begin{equation}
  \ratf=(2\alpha M)^{1/3},
  \quad
  \Kii\Big|_{\min}=-\frac{1}{8\,\alpha}
  \ \text{at}\ r=(4\alpha M)^{1/3},
  \label{eq:ratf}
\end{equation}
compressive (as in Schwarzschild) outside $\ratf$ and stretching inside it.

\begin{figure}[t]
  \includegraphics[width=\columnwidth]{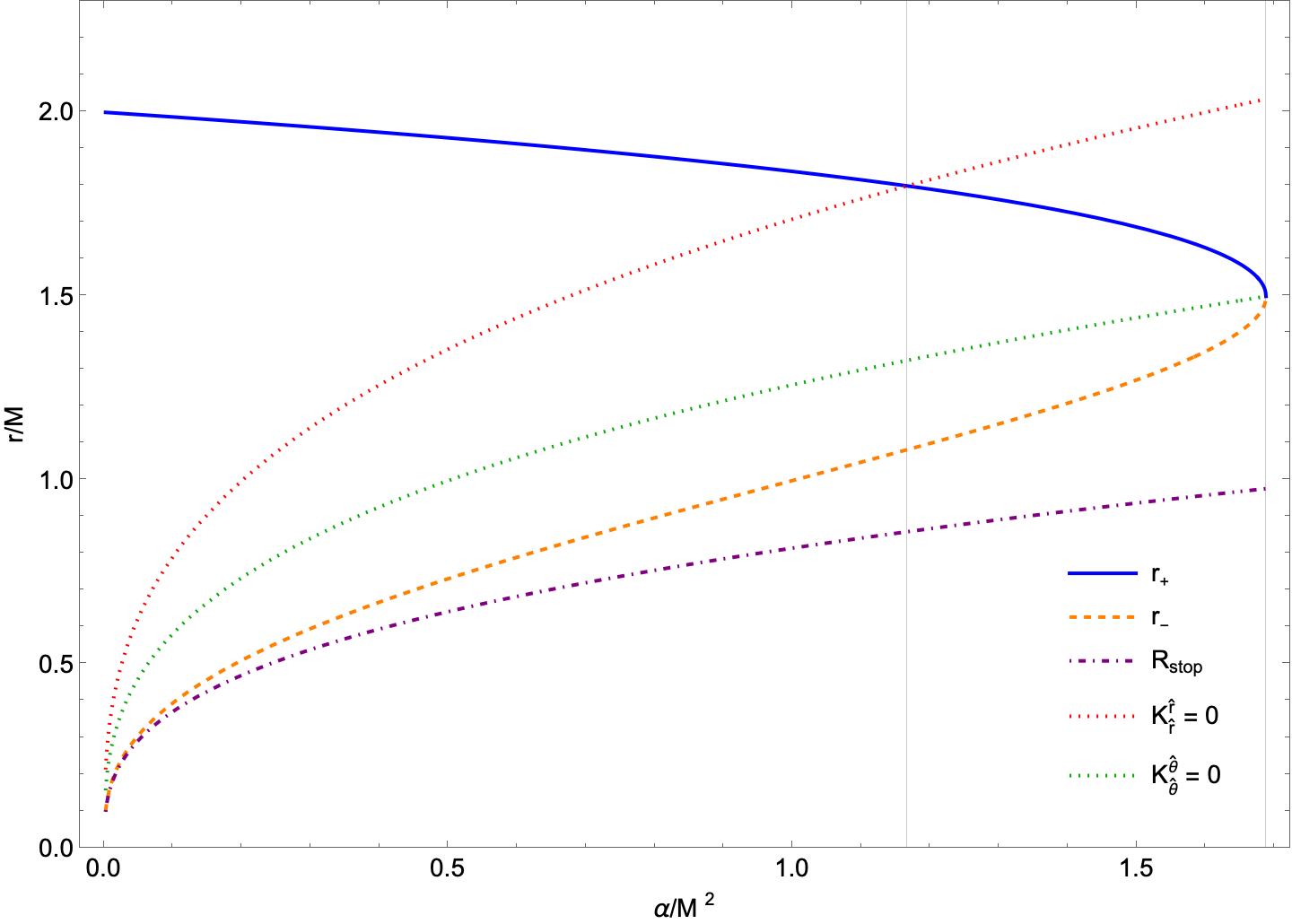}
  \caption{Characteristic radii as functions of $\at=\alpha/M^{2}$: the event
    horizon $x_{+}$, the Cauchy horizon $x_{-}$, the turning point $\rs$ (for
    $b=10M$, a moderate release radius chosen so that its separation from
    $r_{-}$ is visible; Table~\ref{tab:radii} uses $b=100M$, for which the two
    curves would be nearly indistinguishable), and the vanishing-tidal-force
    radii $\rrtf=(5\at)^{1/3}$ and $\ratf=(2\at)^{1/3}$. The angular zero
    remains between the horizons for all $\at$, meeting $x_{+}$ only at
    $\at=27/16$, whereas the radial zero crosses $x_{+}$ at
    $\at_{\star}=729/625$. The turning point stays inside the Cauchy horizon
    throughout.}
  \label{fig:radii}
\end{figure}

Unlike $\rrtf$, this radius cannot escape the horizons, and the reason is
structural. Since $\Kii=-f'/2r$ by Eq.~\eqref{eq:Kgeneral}, the angular
zero is nothing but the stationary point of the metric function; and
$f'(r)=2M/r^{2}-4\alpha M^{2}/r^{5}$ has the single positive root
$r^{3}=2\alpha M$, which is therefore the \emph{global minimum} of $f$, with
$f(\ratf)=1-\tfrac{3}{2}(2\at)^{-1/3}$. Whenever $f$ has two positive roots its
minimum lies between them, so
\begin{equation}
  r_{-}\;<\;\ratf\;\le\;r_{+}
  \qquad\text{for all}\quad 0<\at\le\tfrac{27}{16},
  \label{eq:confinement}
\end{equation}
with equality on the right only at extremality, where $f(\ratf)=0$ and the two
horizons and the angular zero coalesce at $x=3/2$.
\ed{The angular tidal zero therefore remains confined between the horizons
throughout the two-horizon regime.} Figure~\ref{fig:atf} shows the angular
tidal force and Fig.~\ref{fig:radii} displays all of these radii as functions
of $\at$.

\begin{table}[b]
\caption{The tidal sector of the qOS black hole in analytic form. Every radius is
a cube root of $\alpha M$, so every ratio between them is a pure number:
$\ratf/\rs=4^{1/3}$, $\rrtf/\ratf=(5/2)^{1/3}$ and $\rrtf/\rs=10^{1/3}$, with
the ordering $\rs<r_{-}<\ratf<\rrtf$ and $\ratf\le r_{+}$ holding for every
admissible $\at$. The values at the bounce are those of the marginally bound
trajectory, $E=1$.}
\label{tab:summary}
\begin{ruledtabular}
\begin{tabular}{lll}
quantity & value & location \\
\hline
radial tidal zero $\rrtf$   & ---              & $(5\alpha M)^{1/3}$ \\
maximum stretching          & $+1/10\alpha$    & $(10\alpha M)^{1/3}$ \\
angular tidal zero $\ratf$  & ---              & $(2\alpha M)^{1/3}$ \\
maximum transverse squeeze  & $-1/8\alpha$     & $(4\alpha M)^{1/3}$ \\
turning point $\rs$         & ---              & $(\alpha M/2)^{1/3}$ \\
$\Krr$ at the bounce        & $-36/\alpha$     & $\rs$ \\
$\Kii$ at the bounce        & $+6/\alpha$      & $\rs$ \\
anisotropy at the bounce    & $-6$             & $\rs$ \\
peak of $\eta^{\hat r}$     & ---              & $\ratf$ \\
\end{tabular}
\end{ruledtabular}
\end{table}

\subsection{Tidal structure along the infall trajectory}
\label{sec:structure}

The two transitions are \ed{related by a fixed ratio},
\begin{equation}
  \frac{\rrtf}{\ratf}=\Big(\frac{5}{2}\Big)^{1/3}\simeq1.3572,
  \label{eq:ratio}
\end{equation}
independently of $\alpha$ and $M$, with the radial transition always further
out. With Eq.~\eqref{eq:rstopinf} this fixes the whole geometry of the infall,
\begin{equation}
  \rs<\ratf<\rrtf,
  \qquad
  \frac{\ratf}{\rs}=4^{1/3},
  \quad
  \frac{\rrtf}{\rs}=10^{1/3},
  \label{eq:ordering}
\end{equation}
again with parameter-independent ratios. The infalling body therefore always
crosses both transitions, in that order, before it bounces.

At the instant the body comes to rest the tidal components take particularly
simple values. Evaluating them at $\rs=(\alpha M/2)^{1/3}$,
\begin{equation}
  \Krr\big|_{\rs}=-\frac{36}{\alpha},
  \ \
  \Kii\big|_{\rs}=+\frac{6}{\alpha},
  \ \
  \frac{\Krr}{\Kii}\bigg|_{\rs}=-6:
  \label{eq:atturning}
\end{equation}
the body is squeezed radially and stretched transversally, and the ratio is a
pure number independent of $\alpha$ and $M$. These values belong to the
marginally bound trajectory.

For a general turning point, define
$u\equiv\alpha M/r_{\mathrm{stop}}^{3}$. The tidal components at the
turning point are
\begin{align}
  \Krr\big|_{\rs}&=\frac{2M}{\rs^{3}}\,(1-5u),
  \label{eq:generalE}\\[2pt]
  \Kii\big|_{\rs}&=-\frac{M}{\rs^{3}}\,(1-2u),
  \label{eq:generalEb}\\[2pt]
  \frac{\Krr}{\Kii}\bigg|_{\rs}&=-\,\frac{2\,(1-5u)}{1-2u}.
  \label{eq:ratioE}
\end{align}
For the marginally bound trajectory $u=2$, and Eq.~\eqref{eq:ratioE} recovers
$\Krr/\Kii=-6$.

\section{Geodesic deviation}
\label{sec:deviation}

\subsection{Deviation equations and initial conditions}
\label{sec:devsol}

To follow the deformation of an extended body we must integrate
Eq.~\eqref{eq:deviation} rather than merely evaluate the tidal tensor. Using
$dr/d\tau=-\sqrt{E^{2}-f}$ to trade proper time for the radial coordinate,
Eqs.~\eqref{eq:Krr} and~\eqref{eq:Kaa} become linear second-order ODEs in $r$,
\begin{align}
  (E^{2}-f)\,\eta^{\hat r\,\prime\prime}
  -\frac{f'}{2}\,\eta^{\hat r\,\prime}
  +\frac{f''}{2}\,\eta^{\hat r}&=0,
  \label{eq:odeR}\\[2pt]
  (E^{2}-f)\,\eta^{\hat\imath\,\prime\prime}
  -\frac{f'}{2}\,\eta^{\hat\imath\,\prime}
  +\frac{f'}{2r}\,\eta^{\hat\imath}&=0,
  \label{eq:odeA}
\end{align}
where primes denote $d/dr$. These admit the closed-form general solutions
\begin{align}
  \eta^{\hat r}(r)&=\sqrt{E^{2}-f}\,
    \left[C_{1}+C_{2}\!\int\!\frac{dr}{(E^{2}-f)^{3/2}}\right],
  \label{eq:solR}\\[2pt]
  \eta^{\hat\imath}(r)&=r
    \left[C_{3}+C_{4}\!\int\!\frac{dr}{r^{2}\sqrt{E^{2}-f}}\right].
  \label{eq:solA}
\end{align}
That $\eta^{\hat r}=\sqrt{E^{2}-f}=-\dot r$ and $\eta^{\hat\imath}=r$ are
solutions can be verified directly and holds for any $f$; the second solution
in each pair follows by reduction of order.

Following the standard treatment~\cite{CrispinoRN,HaywardTidal} we consider two
physically distinct sets of initial conditions at $r=b>r_{+}$,
\begin{align}
  \eta^{\hat\mu}(b)>0,\quad \dot\eta^{\hat\mu}(b)=0 &\qquad \text{(ICI)},
  \label{eq:ICI}\\
  \eta^{\hat\mu}(b)=0,\quad \dot\eta^{\hat\mu}(b)>0 &\qquad \text{(ICII)}.
  \label{eq:ICII}
\end{align}
ICI describes a cloud of dust particles released from rest at $r=b$ with no
internal velocity dispersion; ICII describes a distribution ejected outward
from a point at $r=b$.

Because $E^{2}=f(b)$, the factor $E^{2}-f$ vanishes at the release radius and
the integrals in Eqs.~\eqref{eq:solR} and~\eqref{eq:solA} require care there.
Two of the four cases nevertheless collapse to elementary analytic forms. Since
$\eta_{1}=\sqrt{E^{2}-f}$ satisfies $\eta_{1}(b)=0$ and
$\dot\eta_{1}(b)=f'(b)/2$, it is by itself the ICII radial solution,
\begin{equation}
  \eta^{\hat r}_{\mathrm{ICII}}(r)
   =\frac{2\,\dot\eta_{0}}{f'(b)}\sqrt{f(b)-f(r)},
  \label{eq:ICIIradial}
\end{equation}
while $\eta^{\hat\imath}=r$ has $\dot\eta=\dot r$, which vanishes at $r=b$, so
\begin{equation}
  \eta^{\hat\imath}_{\mathrm{ICI}}(r)=\eta_{0}\,\frac{r}{b}
  \label{eq:ICIangular}
\end{equation}
exactly. Both hold for \emph{any} metric of the form~\eqref{eq:metric}, a fact
we exploit below as a stringent check on the numerics. The remaining ICII
angular case requires the second solution,
\begin{equation}
  \eta^{\hat\imath}_{\mathrm{ICII}}(r)
   =b\,\dot\eta_{0}\,r\!\int_{r}^{b}\!
     \frac{dr'}{r'^{2}\sqrt{f(b)-f(r')}}.
  \label{eq:ICIIangular}
\end{equation}

\ed{The ICI radial case is the only one for which the quadrature in
Eq.~\eqref{eq:solR} is not elementary. The antiderivative diverges at $r=b$,
and after regularization the result carries an apparent pole at $f'(r)=0$,
which for the qOS metric is located precisely at $r=\ratf$; this pole cancels
against a compensating divergence of the integral. Rather than tracking that
cancellation, we use a first integral.} The solutions $\eta_{1}=-\dot r$ and
the ICI solution $\eta^{\hat r}$ of the same linear equation have a constant
Wronskian, and since $\dot\eta_{1}=f'/2$ this reads
\begin{equation}
  -\dot r\,\dot\eta^{\hat r}-\frac{f'(r)}{2}\,\eta^{\hat r}
  =-\frac{f'(b)}{2}\,\eta_{0},
  \label{eq:wronskian}
\end{equation}
the constant being fixed by the initial data~\eqref{eq:ICI} at $r=b$.
Equivalently,
\begin{equation}
  \dot\eta^{\hat r}_{\mathrm{ICI}}
  =\frac{f'(r)\,\eta^{\hat r}_{\mathrm{ICI}}(r)-f'(b)\,\eta_{0}}
        {2\sqrt{f(b)-f(r)}},
  \label{eq:etadot}
\end{equation}
which is manifestly regular at $\ratf$ and reduces the determination of the
extremum of $\eta^{\hat r}_{\mathrm{ICI}}$ to an algebraic condition.

For the numerical results we integrate the coupled equations directly in
proper time. This formulation is regular at the horizons and terminates at
the turning point. The numerical solutions reproduce the two analytic
solutions~\eqref{eq:ICIIradial} and~\eqref{eq:ICIangular} to relative
accuracies better than $10^{-8}$. In
Figs.~\ref{fig:radICI}--\ref{fig:angICII} the upper panel varies $\at$ at fixed
$b=100M$ and includes the Schwarzschild comparison, while the lower panel
varies $b$ at fixed $\at=0.9$; insets magnify the region $r<4M$, where all of
the structure lies.

\subsection{Radial component}
\label{sec:radialcomp}

\begin{figure}[t]
  \includegraphics[width=\columnwidth]{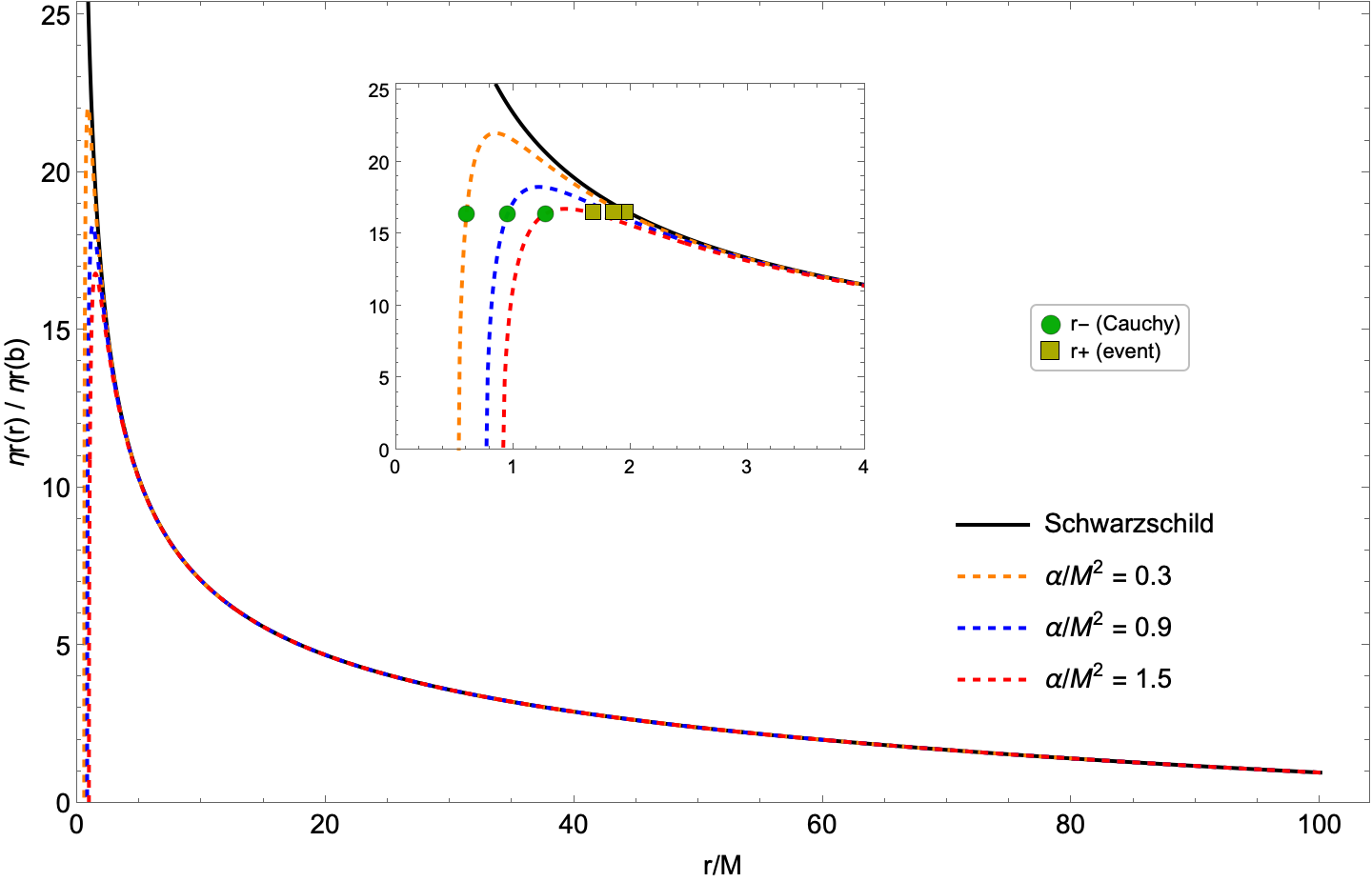}\\[4pt]
  \includegraphics[width=\columnwidth]{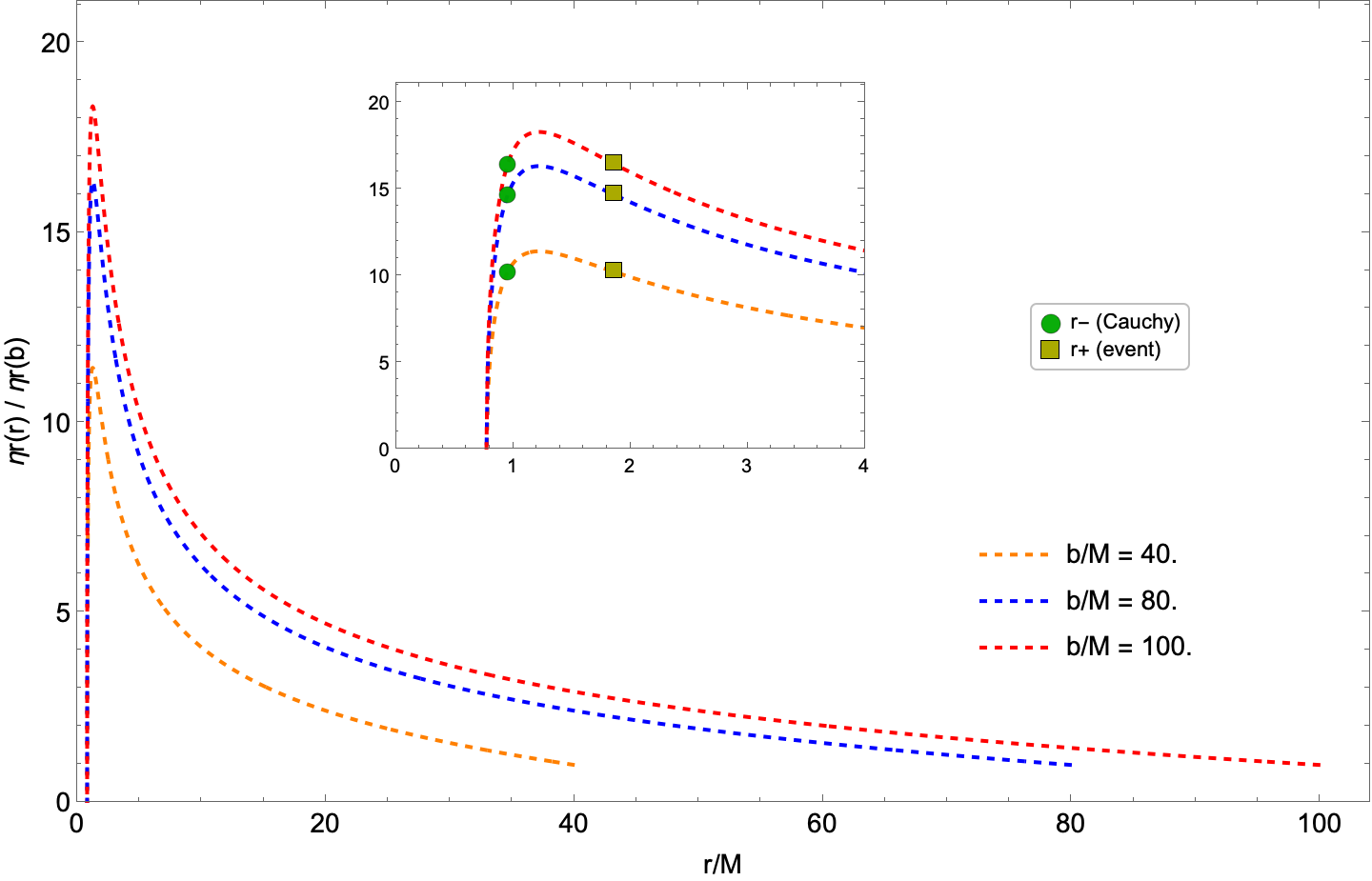}
  \caption{Radial component of the geodesic deviation vector with initial
    condition ICI, \ed{normalized} to its value at $r=b$. Upper panel: several
    $\at$ at $b=100M$, together with the Schwarzschild curve, which diverges at
    the singularity. Lower panel: several release radii $b$ at $\at=0.9$. The
    maximum occurs at the radius fixed by Eq.~\eqref{eq:ICIpeak}, which for
    $b\gg M$ is $\ratf=(2\alpha M)^{1/3}$, the minimum of $f$, to within the
    offset~\eqref{eq:offset} (invisible on this scale; see
    Table~\ref{tab:peak}), and therefore always inside the event horizon;
    thereafter the body is compressed down to the turning point, where
    $\eta^{\hat r}$ is finite.}
  \label{fig:radICI}
\end{figure}

For ICI (Fig.~\ref{fig:radICI}) the radial separation grows during the
exterior infall, continues to grow past the event horizon, \ed{reaches a
maximum, and is then compressed back down to $\rs$, where it remains finite
and is numerically small.} In Schwarzschild the same curve instead diverges at
the singularity.

The peak does not sit at the sign change of $\Krr$, where the radial force
turns compressive, but at the radius fixed by Eq.~\eqref{eq:etadot}. Setting
$\dot\eta^{\hat r}_{\mathrm{ICI}}=0$ gives the exact stationarity condition
\begin{equation}
  f'(r_{\mathrm{peak}})\,\eta^{\hat r}_{\mathrm{ICI}}(r_{\mathrm{peak}})
  =f'(b)\,\eta_{0},
  \label{eq:ICIpeak}
\end{equation}
whose right-hand side is strictly positive for any finite $b>r_{+}$. The peak
therefore lies slightly \emph{outside} $\ratf$, where $f'>0$, rather than
exactly at it. Linearizing $f'$ about the minimum of $f$ and using
$f''(\ratf)=3/\alpha$,
\begin{equation}
  \Delta r\equiv r_{\mathrm{peak}}-\ratf
  \simeq\frac{f'(b)\,\eta_{0}}
             {f''(\ratf)\,\eta^{\hat r}_{\mathrm{ICI}}(r_{\mathrm{peak}})}
  =\frac{\alpha\,f'(b)\,\eta_{0}}{3\,\eta^{\hat r}_{\mathrm{ICI}}(r_{\mathrm{peak}})}.
  \label{eq:offset}
\end{equation}
Since $f'(b)\simeq2M/b^{2}$, the offset vanishes as $b\to\infty$, so that the
statement that $\eta^{\hat r}_{\mathrm{ICI}}$ is maximal at $\ratf=(2\alpha
M)^{1/3}$ survives as an asymptotic identity in the regime of physical
interest. Table~\ref{tab:peak} confirms the offset and its scaling:
the measured displacement matches Eq.~\eqref{eq:offset} to the digits quoted
and drops by two orders of magnitude when $b$ goes from $100M$ to $1000M$.

One might instead expect the peak at the sign change of $\Krr$, that is at
$\rrtf=(5\at)^{1/3}M=1.145M,\,1.651M,\,1.957M$ for $\at=0.3,\,0.9,\,1.5$;
\ed{the measured peaks do not coincide with these values}. Because $\ratf$ lies
between the horizons for every admissible $\at$ by
Eq.~\eqref{eq:confinement}, the maximum radial stretching is always reached
inside the event horizon.

\begin{table}[t]
\caption{Location of the ICI radial-deviation maximum for several
$\at$ and release radii. The measured offset from $\ratf$ agrees with
Eq.~\eqref{eq:offset}.}
\label{tab:peak}
\begin{ruledtabular}
\begin{tabular}{cccccc}
$\at$ & $b/M$ & $r_{\mathrm{peak}}$ & $\ratf$
      & $\Delta r$ & Eq.~\eqref{eq:offset} \\
\hline
0.3 & 100  & 0.843433571 & 0.843432665 & $9.05\times10^{-7}$ & $9.05\times10^{-7}$ \\
0.3 & 1000 & 0.843432668 & 0.843432665 & $2.85\times10^{-9}$ & $2.85\times10^{-9}$ \\
0.9 & 100  & 1.216443669 & 1.216440399 & $3.27\times10^{-6}$ & $3.27\times10^{-6}$ \\
0.9 & 1000 & 1.216440409 & 1.216440399 & $1.03\times10^{-8}$ & $1.03\times10^{-8}$ \\
1.5 & 100  & 1.442255513 & 1.442249570 & $5.94\times10^{-6}$ & $5.94\times10^{-6}$ \\
1.5 & 1000 & 1.442249589 & 1.442249570 & $1.86\times10^{-8}$ & $1.86\times10^{-8}$ \\
\end{tabular}
\end{ruledtabular}
\end{table}

\begin{figure}[t]
  \includegraphics[width=\columnwidth]{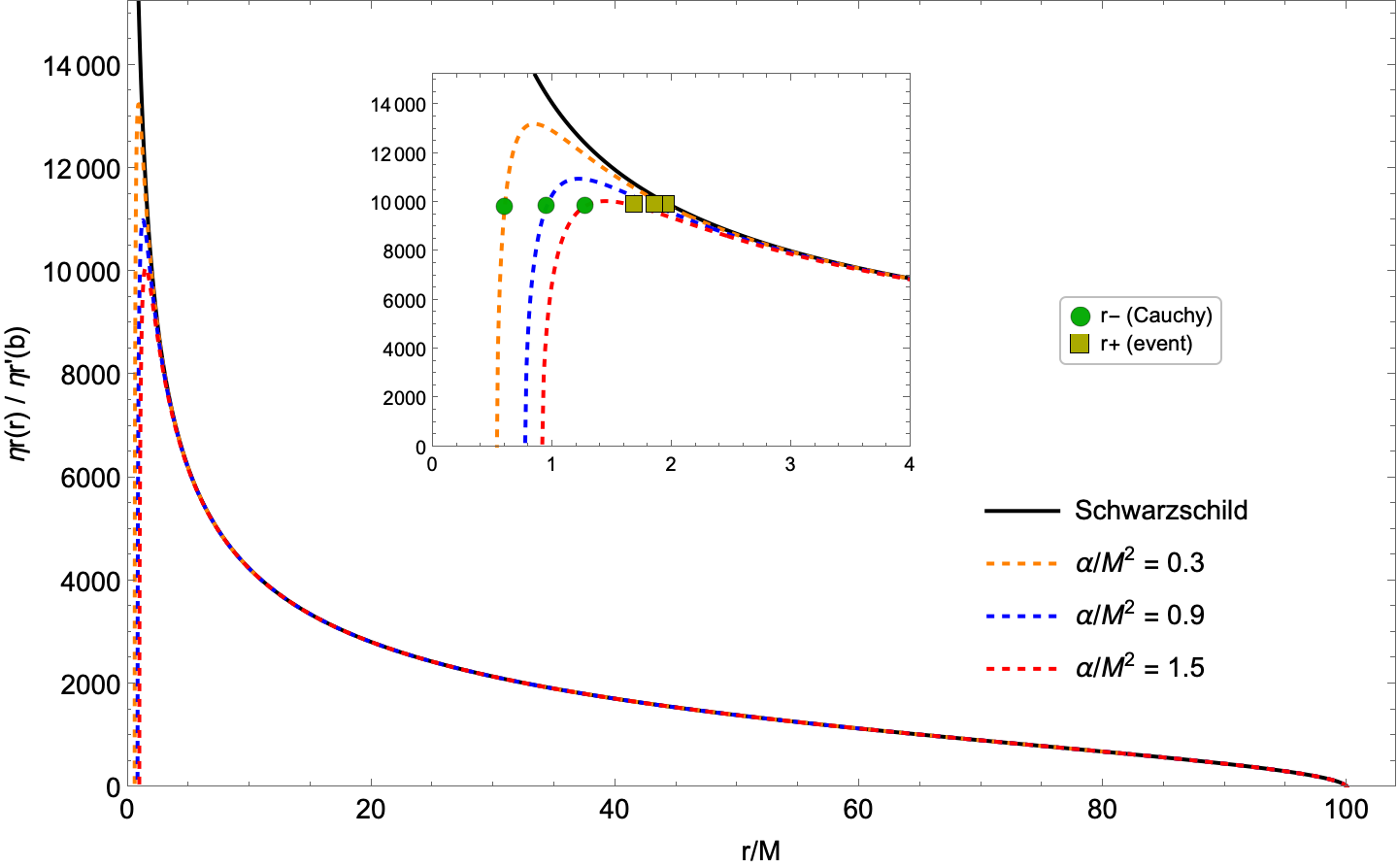}\\[4pt]
  \includegraphics[width=\columnwidth]{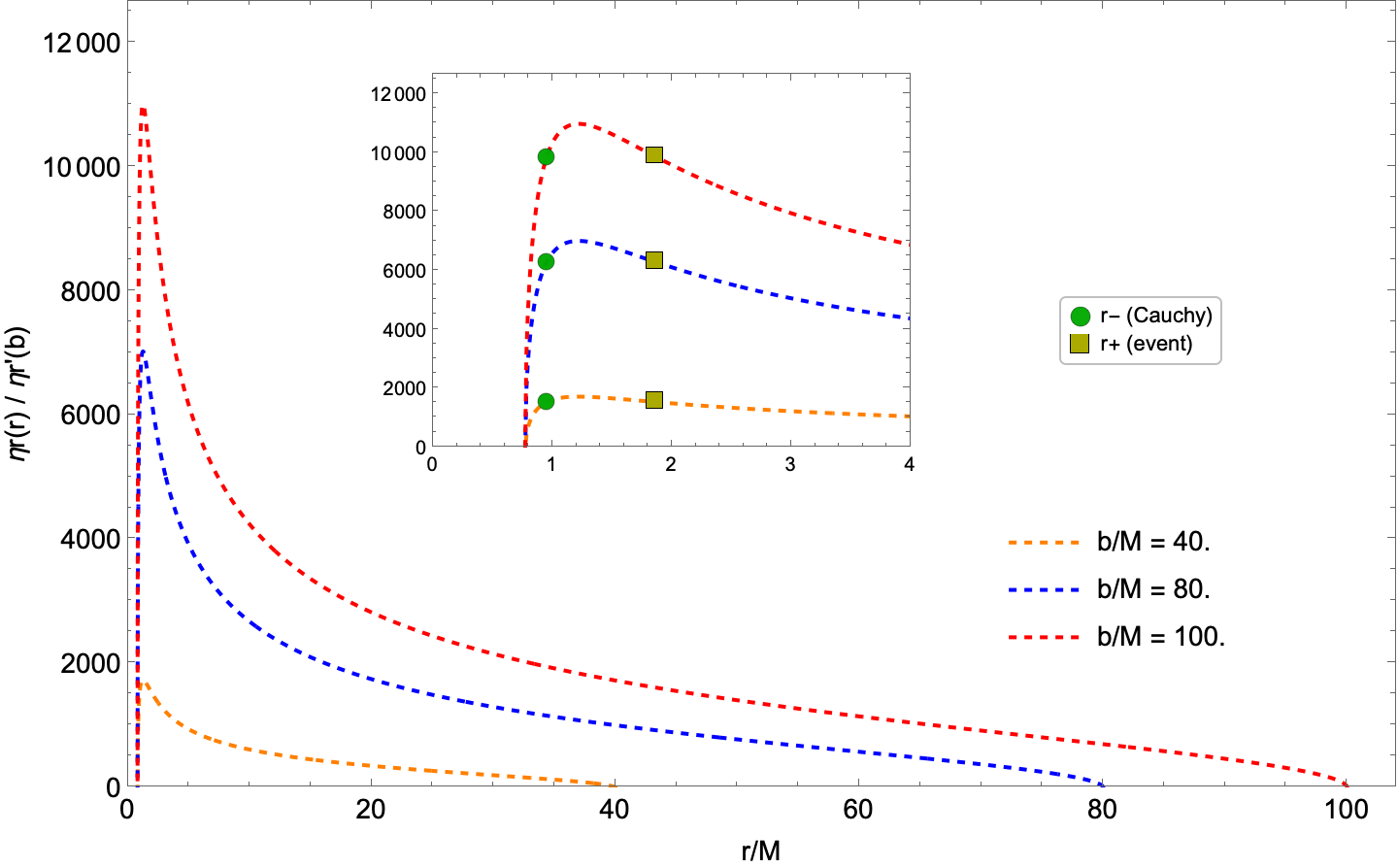}
  \caption{As in Fig.~\ref{fig:radICI}, now for the initial condition ICII. The
    curves follow the exact solution~\eqref{eq:ICIIradial}: they peak at
    $r=\ratf$, where $f$ is minimized, and return exactly to zero at the
    turning point $\rs$. Here the peak is exact at every $b$, in contrast with
    the ICI case. Since ICII has $\eta^{\hat r}(b)=0$, the ordinate is
    \ed{normalized} to $M\dot\eta_{0}$ rather than to $\eta(b)$.}
  \label{fig:radICII}
\end{figure}

For ICII (Fig.~\ref{fig:radICII}), the result follows directly from the exact
solution~\eqref{eq:ICIIradial}. Its derivative is proportional to $-f'$, so the
maximum occurs at $f'(r)=0$. For the metric~\eqref{eq:metric}, this gives
$r=\ratf$ exactly, independently of the release radius $b$. Thus the ICII radial
separation reaches its maximum at $\ratf$, whereas the ICI maximum approaches
$\ratf$ only in the limit $b\to\infty$. The relation
$\Kii=-f'/2r$ from Eq.~\eqref{eq:Kgeneral} shows that the ICII maximum
coincides with the zero of the angular tidal component. This result holds for
any metric of the form~\eqref{eq:metric}, not only for qOS.

At the turning point, $f(\rs)=f(b)$ by definition, and hence
$\eta^{\hat r}_{\mathrm{ICII}}(\rs)=0$. A radial distribution initially
released from a point at $r=b$ therefore has vanishing radial separation again
at $r=\rs$.

The ICI case does the same, but only asymptotically. Evaluating the
Wronskian~\eqref{eq:wronskian} at the turning point, where $\dot r=0$, gives
the radial deviation at the bounce in analytic form,
\begin{equation}
  \eta^{\hat r}_{\mathrm{ICI}}(\rs)=\frac{f'(b)}{f'(\rs)}\,\eta_{0},
  \label{eq:etaatbounce}
\end{equation}
valid for any static spherically symmetric metric possessing a turning point;
we have verified it to a relative accuracy of $10^{-8}$ for $\at=0.9$ and $2.5$
at $b/M=50$ and $100$. Since $f'(\rs)<0<f'(b)$, the value is negative and of
order $M/b^{2}$, so the radial deviation vector stays finite at the bounce and
in fact tends to zero as $b\to\infty$. In the marginally bound limit both
sets of initial conditions therefore refocus the radial extent of the body to a
point at $\rs$: ICII exactly and at any $b$, ICI asymptotically. \ed{The sign
is also meaningful: the radial separation reverses immediately before the
bounce, although the effect is of order $M/b^{2}$ and vanishes for a distant
release.}

\subsection{Angular component}
\label{sec:angularcomp}

\begin{figure}[t]
  \includegraphics[width=\columnwidth]{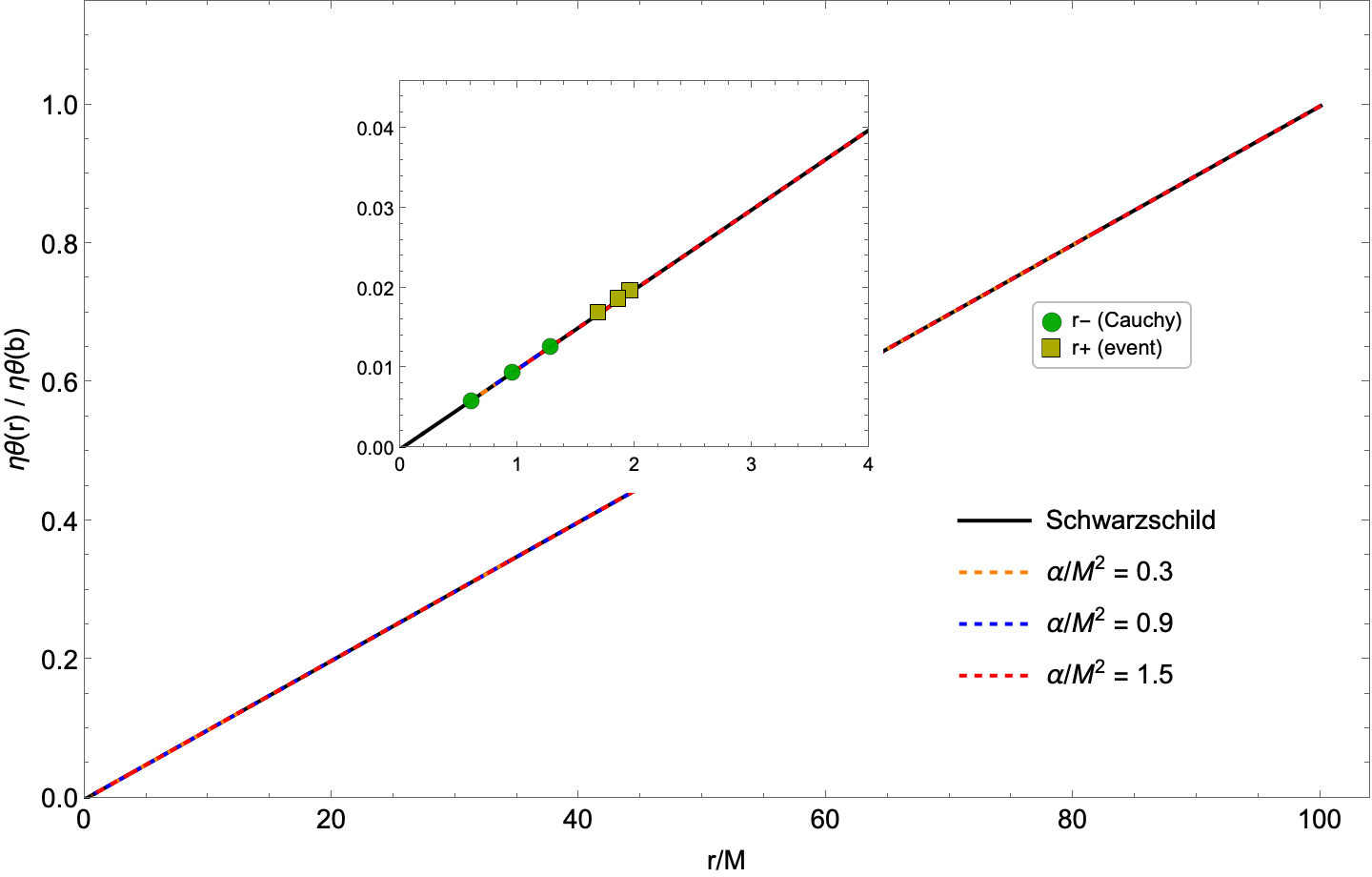}\\[4pt]
  \includegraphics[width=\columnwidth]{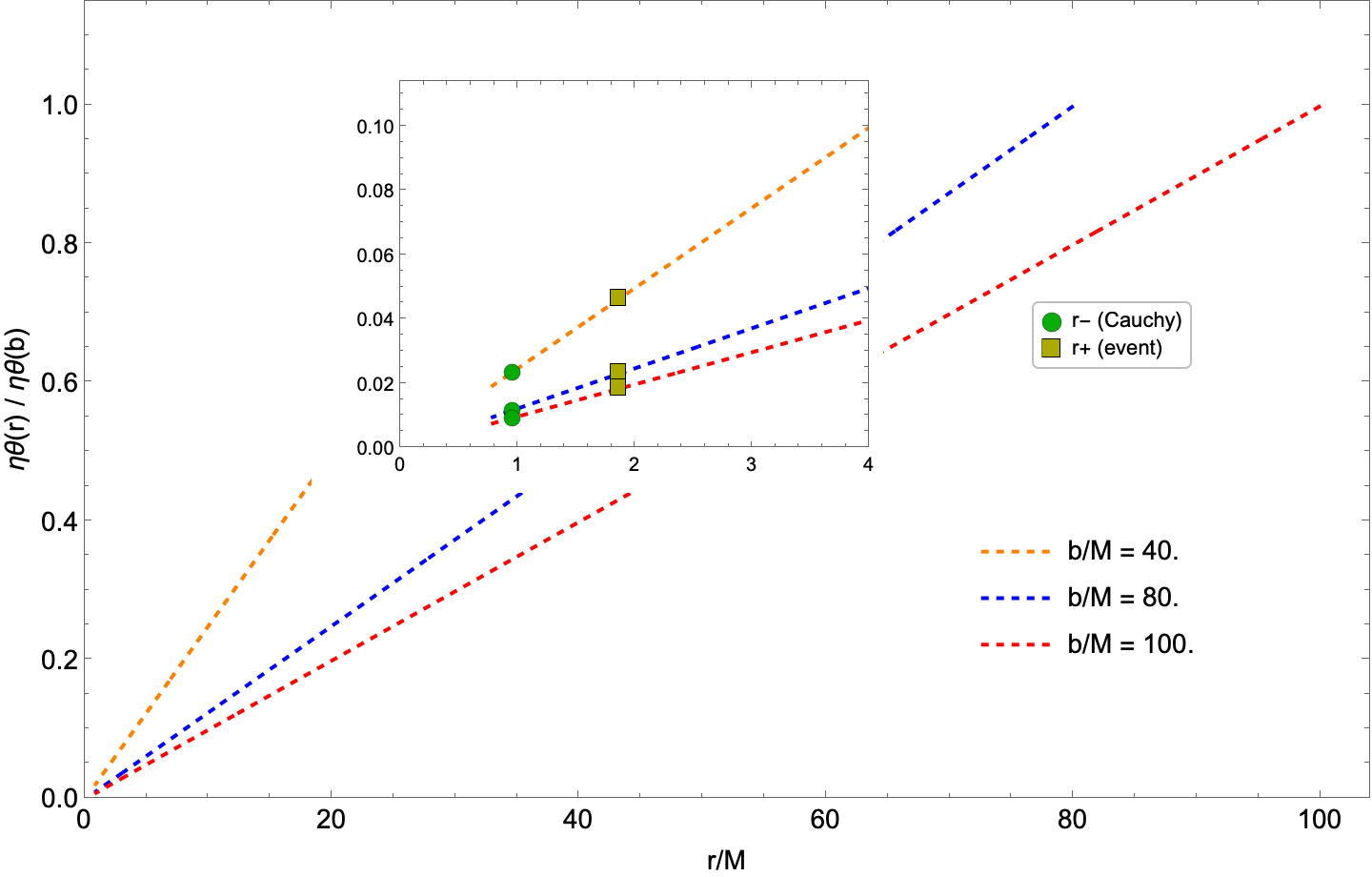}
  \caption{Angular component of the geodesic deviation vector with initial
    condition ICI. The \ed{behavior} is the exact linear law
    $\eta^{\hat\imath}=\eta_{0}r/b$ of Eq.~\eqref{eq:ICIangular}, which is
    independent of the metric function; the curves for different $\at$ and for
    Schwarzschild therefore coincide in the upper panel, and in the lower panel
    differ only through the slope $1/b$.}
  \label{fig:angICI}
\end{figure}

For ICI (Fig.~\ref{fig:angICI}), the angular deviation is given exactly by
\begin{equation}
\eta^{\hat\imath}=\eta_{0}\,r/b,
\end{equation}
independently of the metric function. The transverse separation therefore
decreases linearly with the areal radius and remains finite at the turning
point.

\begin{figure}[t]
  \includegraphics[width=\columnwidth]{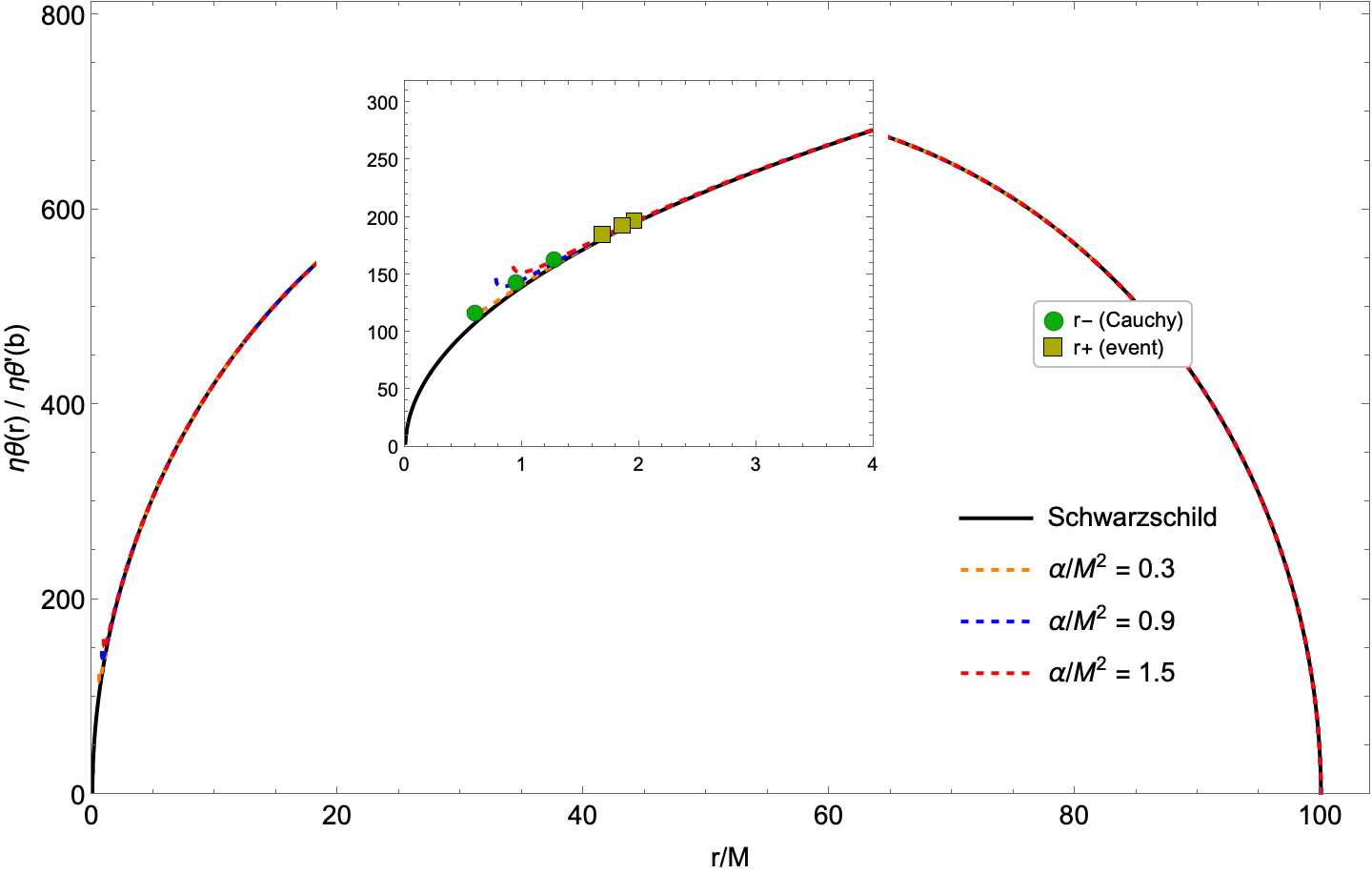}\\[4pt]
  \includegraphics[width=\columnwidth]{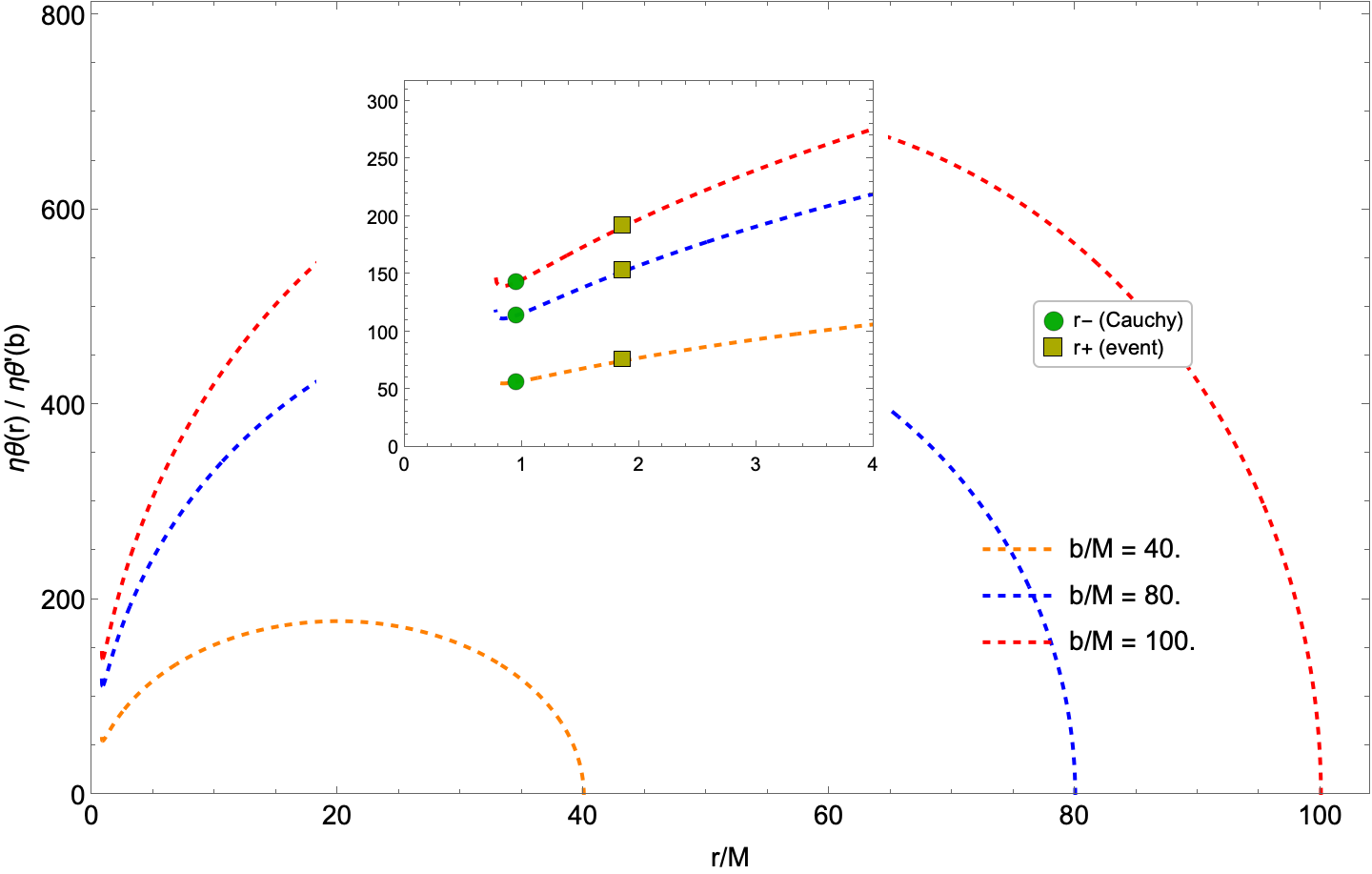}
  \caption{As in Fig.~\ref{fig:angICI}, now for ICII. The transverse separation
    rises to a maximum, falls, and rises again inside the event horizon\ed{:
    the} signature of the sign change of the angular tidal force at
    $\ratf=(2\alpha M)^{1/3}$. As in Fig.~\ref{fig:radICII}, the ordinate is
    \ed{normalized} to $M\dot\eta_{0}$, since $\eta^{\hat\imath}(b)=0$ for
    ICII.}
  \label{fig:angICII}
\end{figure}

For ICII (Fig.~\ref{fig:angICII}) the angular separation first grows,
reaches a maximum roughly halfway along the trajectory, and then decreases as
the body approaches the black hole. Inside the event horizon the curve passes
through a minimum and turns upward again before the turning point. This
non-monotonic \ed{behavior} is the direct imprint of the sign change of the
angular tidal force at $\ratf$: transverse compression outside, transverse
stretching inside.

\section{Origin of the tidal signatures}
\label{sec:origin}

The previous sections identified several characteristic features of the tidal
sector: the two tidal zeros and their ratio $(5/2)^{1/3}$, the extrema at
$1/(10\alpha)$ and $-1/(8\alpha)$, and the bounce anisotropy
$\Krr/\Kii=-6$. We now examine which of these features follow from the
$r^{-4}$ quantum correction and which are more general properties of the
metric.

We first rewrite the tidal tensor in terms of the effective source associated
with the metric. This gives a direct interpretation of the tidal zeros and of
the anisotropy at the bounce. We then vary $\alpha$ to distinguish the role of
its sign from that of its magnitude.

\subsection{The effective source}
\label{sec:source}

The qOS metric does not solve Einstein's equations, since the term
$\alpha M^{2}/r^{4}$ comes from the effective loop dynamics and modifies the
left-hand side of the field equations. \ed{One may nevertheless insert
Eq.~\eqref{eq:metric} into $G_{\mu\nu}=8\pi T_{\mu\nu}$ and \emph{define}
$T_{\mu\nu}$ through the Einstein tensor of the metric. The resulting
$T_{\mu\nu}$ is an effective description rather than the stress tensor of a
physical medium. It provides a convenient translation of the quantum correction
into effective densities and pressures, which can then be related to the tidal
tensor.} The corresponding density and pressures are completely fixed by the
metric. The $tt$ equation $m'=4\pi r^{2}\rho$ fixes $\rho$ from the
Misner--Sharp mass $m=M-\alpha M^{2}/2r^{3}$ of Sec.~\ref{sec:qos}. Since the
metric satisfies $g_{tt}g_{rr}=-1$, one has $G^{t}{}_{t}=G^{r}{}_{r}$ and
therefore $p_{r}=-\rho$. Conservation, $\nabla_{\mu}T^{\mu}{}_{r}=0$,
then gives $p_{t}=-\rho-\tfrac{r}{2}\rho'$. Hence
\begin{equation}
  \rho=-p_{r}=\frac{3\alpha M^{2}}{8\pi r^{6}},
  \qquad
  p_{t}=2\rho,
  \label{eq:source}
\end{equation}
and more generally $\rho\propto r^{-k}$ implies $p_{t}/\rho=k/2-1$, which
gives $p_{t}=\rho$ for the Reissner--Nordstr\"om source
$\rho=Q^{2}/8\pi r^{4}$, as expected for a Maxwell field.

Table~\ref{tab:ec} summarizes the energy conditions. Three of the four are
satisfied. \ed{The only violation is of the dominant energy condition and is
caused by the transverse pressure. Since the spacetime region is vacuum,
$\rho$, $p_{r}$ and $p_{t}$ do not represent a physical material medium, and
the violation therefore characterizes the effective-source representation of
the modified field equations rather than the underlying geometry. It is also
directly responsible for the tidal anisotropy discussed below.}

\begin{table}[b]
\caption{Energy conditions for the effective source~\eqref{eq:source}. The
radial sector saturates every bound because $p_{r}=-\rho$ identically for
metrics with $g_{tt}g_{rr}=-1$; only the transverse sector violates the
dominant condition.}
\label{tab:ec}
\begin{ruledtabular}
\begin{tabular}{llll}
condition & requirement & qOS value & status \\
\hline
NEC & $\rho+p_{i}\ge0$ & $0$ and $3\rho$ & satisfied \\
WEC & NEC\ed{,} $\rho\ge0$ & \ed{$3\alpha M^{2}/8\pi r^{6}$} & satisfied \\
SEC & NEC\ed{,} $\rho+p_{r}+2p_{t}\ge0$ & $4\rho$ & satisfied \\
DEC & WEC\ed{,} $|p_{i}|\le\rho$ & $|p_{t}|=2\rho$ & \textbf{violated} \\
\end{tabular}
\end{ruledtabular}
\end{table}

Using $m'=4\pi r^{2}\rho$ and $r\rho'=-2(\rho+p_{t})$ in
Eq.~\eqref{eq:Kgeneral} gives the exact translation
\begin{equation}
  \Krr = \frac{2m}{r^{3}}-8\pi(\rho+p_{t}),
  \qquad
  \Kii = 4\pi\rho-\frac{m}{r^{3}},
  \label{eq:Ksource}
\end{equation}
valid for any metric of the form~\eqref{eq:metric} and returning the
Schwarzschild values at $\rho=p_{t}=0$. \ed{The transverse pressure thus enters
the radial tidal component through $\rho+p_{t}$, whereas the angular component
depends on the competition between the local density and the enclosed mass.}
The two zeros identified in Sec.~\ref{sec:tidal} therefore have a simple
interpretation: $\Kii=0$ where $4\pi\rho=m/r^{3}$, that is, where the effective
density overtakes the enclosed mass, and $\Krr=0$ where
$4\pi(\rho+p_{t})=m/r^{3}$. Their separation is fixed by the transverse
equation of state alone: for a correction $c/r^{n}$, which
has $p_{t}/\rho=n/2$,
\begin{equation}
  \left(\frac{\rrtf}{\ratf}\right)^{n-1}=\frac{n+1}{2}
  =\frac{1}{2}+\frac{p_{t}}{\rho}.
  \label{eq:radiiratio}
\end{equation}
For $n=4$ this gives $5/2$, while the Reissner--Nordstr\"om case $n=2$ gives
$3/2$, consistent with $r=3Q^{2}/2M$ and $r=Q^{2}/M$~\cite{CrispinoRN}.

The relation becomes particularly transparent at the bounce. At a marginally
bound turning point $m(\rs)=0$, so Eq.~\eqref{eq:Ksource} collapses to
$\Krr=-8\pi(\rho+p_{t})$ and $\Kii=4\pi\rho$, whence
\begin{equation}
  \frac{\Krr}{\Kii}\bigg|_{\rs}=-2\left(1+\frac{p_{t}}{\rho}\right).
  \label{eq:ratioEOS}
\end{equation}
For the qOS solution with $\alpha>0$ one has $\Kii|_{\rs}=4\pi\rho>0$, so a
body is stretched transversally at the moment it comes to rest. \ed{The
dominant energy condition requires $p_{t}\le\rho$, which bounds the ratio by
$\Krr/\Kii\ge-4$, with equality for a Maxwell-type source; this is exactly the
Reissner--Nordstr\"om value. The qOS ratio $-6$ lies strictly outside that
bound, so the qOS bounce can be distinguished from a bounce supported by a
DEC-respecting source through this local tidal observable.} More generally, a
correction $c/r^{n}$ gives $-(n+2)$: the measured anisotropy returns the
exponent of the correction directly.

\subsection{Dependence on the sign and magnitude of $\alpha$}
\label{sec:regimes}

\begin{figure*}[t]
  \includegraphics[width=0.9\textwidth]{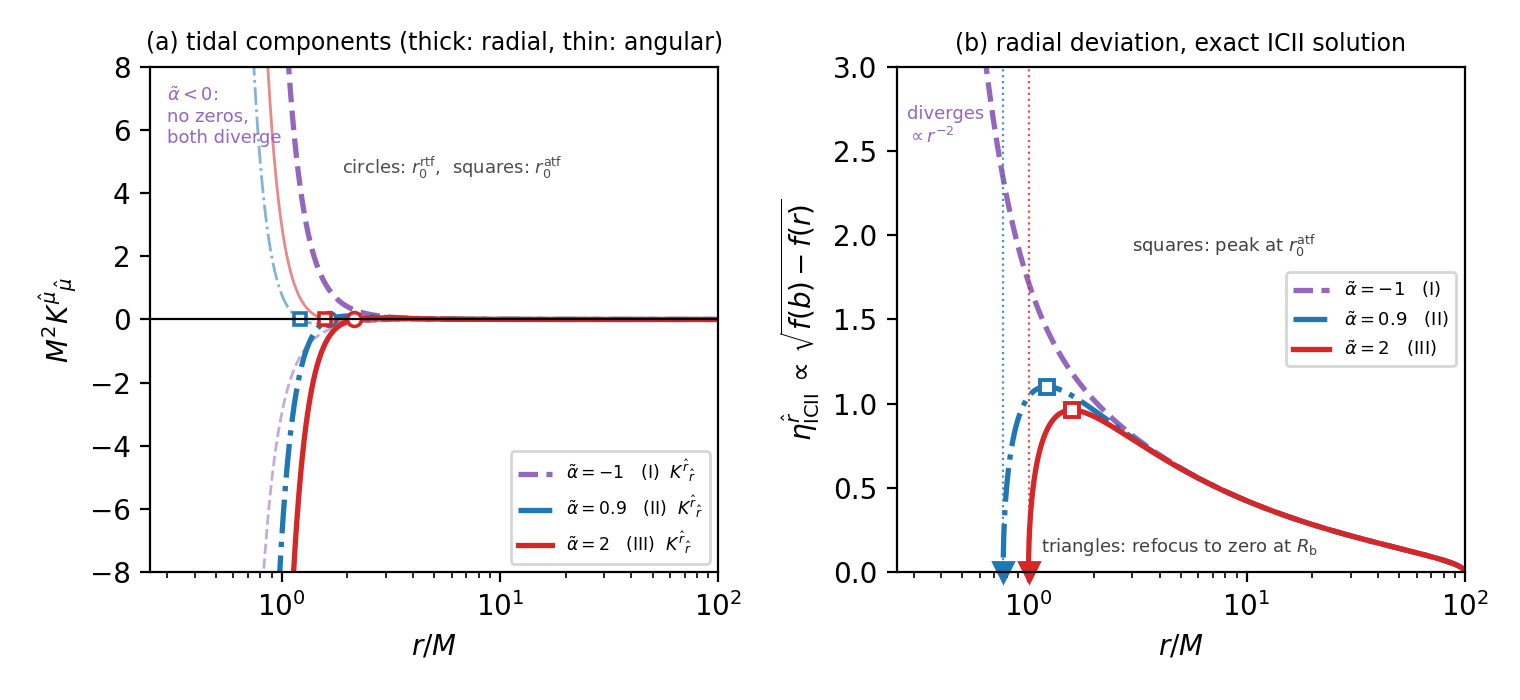}
  \caption{The tidal sector across the three regimes. (a) Tidal components
    (thick: radial; thin: angular). For $\at<0$ neither component has a zero or
    an extremum and both diverge as $r^{-6}$ with the Schwarzschild signs; for
    $\at>0$ the zeros $\rrtf=(5\at)^{1/3}$ (circles) and $\ratf=(2\at)^{1/3}$
    (squares) are present in the two-horizon and the horizonless case alike.
    (b) The exact ICII radial deviation
    $\eta^{\hat r}\propto\sqrt{f(b)-f(r)}$ for $b=100M$: it diverges as
    $r^{-2}$ for $\at<0$, and for $\at>0$ peaks at $\ratf$ and refocuses to
    zero at the turning point $\rs$ (triangles). The $\at=2$ curves have the
    same shape as those of the black-hole window; the difference is that their
    structure lies outside any horizon.}
  \label{fig:devregimes}
\end{figure*}

In the qOS construction $\alpha$ is positive and fixed, and horizons exist only
for $\at\le27/16$. \ed{Stepping outside that window is not a proposal for new
physics; it serves to isolate the assumptions on which the results of
Secs.~\ref{sec:tidal} and~\ref{sec:deviation} rest.} Table~\ref{tab:regimes}
summarizes the comparison and Fig.~\ref{fig:devregimes} displays it.

For $\alpha<0$ the tidal structure changes qualitatively. Both terms of $f'$
are positive and both terms of $f''$ are negative at every radius, so $f$ rises
monotonically from $-\infty$ at the \ed{center} to $1$ at infinity: there is a
single horizon, now \emph{outside} its Schwarzschild value, and a spacelike
singularity, as in Schwarzschild. The zeros $\ratf$ and $\rrtf$, as well as the
extrema at $1/(10\alpha)$ and $-1/(8\alpha)$, are then absent rather than
shifted to different radii. Panel~(a) of Fig.~\ref{fig:devregimes} shows that
both components retain the Schwarzschild signs but diverge more rapidly, as
$r^{-6}$ rather than $r^{-3}$.

Nor is there a bounce, since $m=M+|\alpha|M^{2}/2r^{3}$ never vanishes, so the
body reaches $r=0$; by the exact solution~\eqref{eq:ICIIradial},
$f(b)-f(r)\to|\alpha|M^{2}/r^{4}$ there and the radial deviation diverges as
$\eta^{\hat r}\sim M\sqrt{|\alpha|}/r^{2}$, against $r^{-1/2}$ for
Schwarzschild. This \ed{behavior} corresponds to the steep branch in panel~(b),
where the measured logarithmic slope is $-2.0000$ over
$10^{-3}<r/M<2\times10^{-2}$. Finally, the effective source~\eqref{eq:source}
now has $\rho<0$, so the null and weak energy conditions fail. Negative
$\alpha$ therefore demands exotic matter and leads to more severe
spaghettification than in the Schwarzschild case, with no bounce to cut it off.
It is thus the positive sign of the loop-quantum correction that produces the
bounded, sign-changing tidal sector.

\ed{The magnitude of $\alpha$ plays a different role: it does not change the
tidal sector itself, but determines whether that structure is hidden behind the
horizon or exposed outside it.} Because $\ratf$ is the global minimum of $f$
(Sec.~\ref{sec:angulartidal}), the horizon structure is fixed by the sign of
$f$ at the \emph{angular tidal zero}, $f(\ratf)=1-\tfrac{3}{2}(2\at)^{-1/3}$,
which is negative for $\at<27/16$, zero at extremality and positive beyond. For
$\at>27/16$ one has $f>0$ everywhere, and although the singularity is naked,
massive radial probes still turn around at $\rs$, whereas radial null geodesics
reach $r=0$ in finite affine parameter.

The tidal expressions are unchanged, and so are the two zeros, their ordering
in Eq.~\eqref{eq:ordering}, the maximum of $\eta^{\hat r}$ at $\ratf$, the
refocusing at $\rs$ and the ratio $-6$. For $\at=2$ and $b=100M$ the
integration gives $\rs=1.0034M$ and, at the bounce, $\Krr=-17.62/M^{2}$ and
$\Kii=+2.93/M^{2}$, approaching $-18/M^{2}$ and $+3/M^{2}$ as $b\to\infty$. A
distant observer could therefore follow the whole deformation history,
refocusing included, whereas for $\at<27/16$ the same statement concerns the
black-hole interior. This is qualitatively similar to the naked-singularity
branch of the $q$-metric, where \ed{the} coordinate singularity at $r=2m$ is
likewise approached by a deviation vector that is stretched or compressed
according to the deformation parameter~\cite{Idrissov2025}, the difference
being that here the body is turned back before reaching it.

\begin{figure}[t]
  \includegraphics[width=\columnwidth]{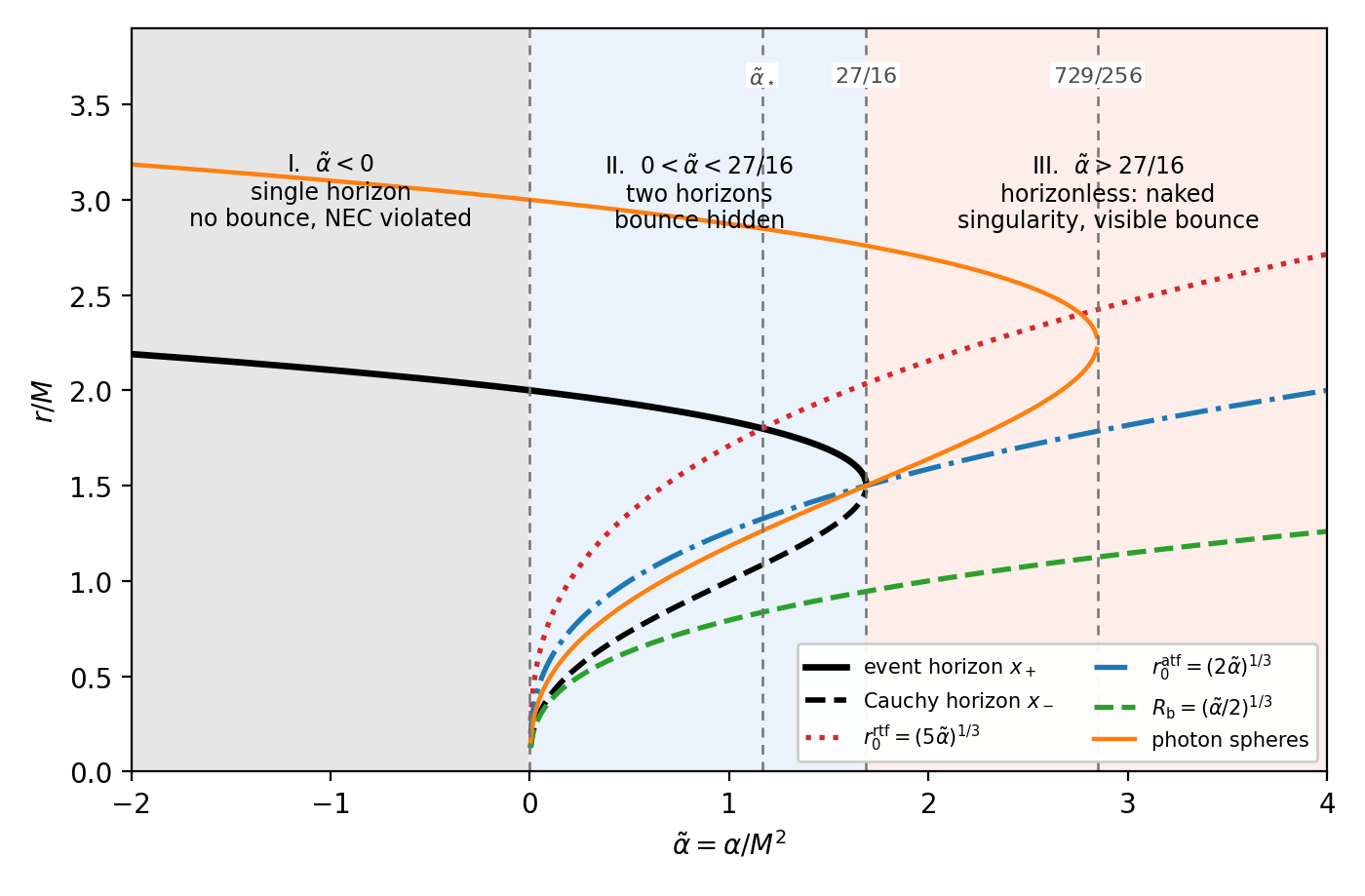}
  \caption{Characteristic radii over the full range of $\at=\alpha/M^{2}$.
    Region~I ($\at<0$): a single horizon $x_{+}>2$ and no other characteristic
    radius\ed{, so that there are} no tidal zeros and no turning point.
    Region~II ($0<\at<27/16$): the black-hole window of Secs.~\ref{sec:tidal}
    and~\ref{sec:deviation}, with the radial tidal zero crossing $x_{+}$ at
    $\at_{\star}=729/625$. Region~III ($\at>27/16$): no horizon, and all three
    characteristic radii exposed in the exterior in the same order and with the
    same ratios. The photon spheres merge and disappear at $\at=729/256$; at
    $\at=27/16$ the degenerate horizon, $\ratf$ and the inner photon sphere
    meet at $x=3/2$.}
  \label{fig:regimes}
\end{figure}

\ed{Turning to the null sector}, circular photon orbits satisfy $rf'-2f=0$,
that is, $\at=\tfrac{1}{3}x^{3}(3-x)$, which peaks at $x=9/4$ with
$\at_{\mathrm{ph}}=729/256$. For $27/16<\at<729/256$ the spacetime is
horizonless and ultracompact, with two light rings, of which the inner one is
stable. For $\at>729/256$ no circular null orbits exist and hence there is no
shadow. At extremality the degenerate horizon, the angular tidal zero and the
inner photon sphere all meet at $x=3/2$. Figure~\ref{fig:regimes} collects the
characteristic radii across the three regions.

These extensions beyond the physical parameter range are used only to
characterize the mathematical structure of the solution, not as predictions of
the underlying quantum theory. The regime $\at>27/16$ corresponds to
$M\lesssim M_{\mathrm{Pl}}$, while $\at<0$ lies outside the loop-quantization
construction.

\begin{table*}
\caption{What the tidal sector depends on. Region~II is the black-hole window
\ed{analyzed} in Secs.~\ref{sec:tidal} and~\ref{sec:deviation}. The sign of the
quantum correction decides whether the tidal sector has any structure; its
magnitude decides only whether that structure lies inside the event horizon.}
\label{tab:regimes}
\begin{ruledtabular}
\begin{tabular}{lccc}
 & I: $\at<0$ & II: $0<\at<27/16$ & III: $\at>27/16$ \\
\hline
horizons & one, $x_{+}>2$ & two, $x_{-}<x_{+}$ & none \\
singularity & spacelike, hidden & hidden & naked, photons only \\
tidal zeros $\ratf,\rrtf$ & neither exists & interior & exterior \\
bounds $1/10\alpha,\,-1/8\alpha$ & absent & attained, hidden & attained, exposed \\
tidal signs as $r\to0$ & Schwarzschild, $\propto r^{-6}$ & reversed & reversed \\
turning point $\rs$ & none & inside $r_{-}$ & exterior \\
bounce ratio $\Krr/\Kii$ & --- & $-6$ & $-6$ \\
inner $\eta^{\hat r}$ & diverges as $r^{-2}$ & refocuses at $\rs$ & refocuses at $\rs$ \\
energy conditions & NEC, WEC violated & only DEC violated & only DEC violated \\
\end{tabular}
\end{ruledtabular}
\end{table*}

\section{Conclusions}
\label{sec:conclusion}

We have \ed{analyzed} the tidal sector of the quantum Oppenheimer--Snyder black
hole. In a frame adapted to radial free fall, the tidal tensor approaches the
Schwarzschild form at large radii but changes sign in the inner region. The
characteristic radii and bounds obtained in the analysis are summarized in
Table~\ref{tab:summary}.

The angular tidal component vanishes at
$\ratf=(2\alpha M)^{1/3}$. For the admissible two-horizon range,
$r_{-}<\ratf\le r_{+}$, with equality at extremality. The radial component
instead vanishes at $\rrtf=(5\alpha M)^{1/3}$ and crosses the event horizon at
$\at=729/625$. Thus, for sufficiently large $\at$, the radial
stretching-to-compression transition occurs outside the event horizon.

The geodesic deviation analysis shows that the extrema of the separation do
not in general coincide with the zeros of the corresponding tidal components.
For ICII, the radial separation reaches its maximum exactly at $\ratf$ for any
release radius $b$. For ICI, the maximum lies slightly above $\ratf$ at finite
$b$ and approaches $\ratf$ as $b\to\infty$. This follows from
$\Kii=-f'/2r$ and therefore holds for any metric of the form
\eqref{eq:metric}. The radial separation vanishes again at the turning point
for ICII and approaches zero there for ICI in the distant-release limit. Both
remain finite along the accessible trajectory, in contrast with Schwarzschild.

For ICI, the angular deviation has the metric-independent form
$\eta^{\hat\imath}=\eta_{0}r/b$. The ICII angular component is non-monotonic
and is controlled by the sign change of $\Kii$. At the turning point of a
marginally bound trajectory,
\begin{equation}
\Krr=-\frac{36}{\alpha},\qquad
\Kii=\frac{6}{\alpha},\qquad
\frac{\Krr}{\Kii}=-6.
\label{eq:concl}
\end{equation}
More generally, the two tidal zeros are related to the effective energy density
and enclosed mass, while their ratio is fixed by the transverse equation of
state. The bounce anisotropy is
$-2(1+p_{t}/\rho)$ and is bounded below by $-4$ for sources satisfying the
dominant energy condition. The qOS value $-6$ therefore requires a violation
of the dominant energy condition in the effective-source description.

The sign of $\alpha$ determines the qualitative structure of the tidal sector.
For $\alpha>0$, the inner repulsive term produces a turning point for radial
timelike geodesics with sufficiently small energy. The magnitude of $\alpha$
sets the location of the tidal structure and determines whether it is hidden
behind the event horizon. For $\alpha<0$, the tidal zeros and turning point are
absent, and the singular \ed{behavior} remains exposed.

The qOS geometry differs from regular black hole models with de~Sitter cores.
In the latter, the curvature and tidal forces remain finite at the
\ed{center}. In qOS, the Kretschmann scalar diverges as $r^{-12}$, despite the
existence of a turning point for radial timelike motion. The finiteness of the
tidal deformation along such trajectories is therefore a consequence of the
accessible domain of the geodesic rather than of curvature regularity. This
protection is not universal: radial null geodesics reach the
\ed{center}, while timelike geodesics with sufficiently large energy can turn
around at arbitrarily small radii.

The analysis also has a limited domain of direct applicability. The qOS
metric is a vacuum exterior matched to the collapsing matter distribution, so
the characteristic radii need not all lie in the vacuum region for a given
collapse history. Moreover, the turning point lies inside the Cauchy horizon,
where mass inflation is expected to affect the classical evolution
\cite{PoissonIsrael1989,PoissonIsrael1990}. A complete treatment would
therefore require the backreaction of this instability and a consistent
description of the interior.
Extending the analysis across the matching surface and to nonradial geodesics
are natural directions for future work.

\begin{acknowledgments}
AI acknowledges financial support from SECIHTI through the National Postgraduate
Scholarship Program (CVU 2222058). This work was supported by DGAPA-PAPIIT UNAM, Grant No. 108225, and
Conahcyt, grant No. CBF-2025-I-243.
\end{acknowledgments}

\bibliographystyle{apsrev4-2}
\bibliography{bib}

\end{document}